\documentclass[a4paper]{article}
\usepackage[style=numeric-comp,  sorting=none, isbn=false]{biblatex}
\usepackage[colorlinks=true, allcolors=blue]{hyperref}
\usepackage[font={small,sl},labelfont=bf]{caption}
\usepackage{placeins}
\usepackage{graphicx}
\usepackage[a4paper,top=3cm,bottom=3cm,left=3cm,right=3cm,marginparwidth=1.75cm]{geometry}
\usepackage[utf8]{inputenc}
\usepackage[colorinlistoftodos, textwidth=4.6cm]{todonotes}
\usepackage{amsmath, amssymb}

\makeatletter
\g@addto@macro\appendix{}
\makeatother

\newcommand{\figref}[2][]{%
\ifx\FirstArg\empty
Figure \ref{fig:#2}%
\else
Figure \ref{fig:#2}\subref{#1}%
\fi}

\newcommand{\subref}[1]{\textbf{#1}}
\newcommand{\subf}[1]{\subref{#1}\textbf{,}}

\newcommand{\eV}{\text{\,eV}}
\title{Imaging moir\'e deformation and dynamics in twisted bilayer graphene}
\author{Tobias A. de Jong$^{1,*}$, Tjerk Benschop$^1$, Xingchen Chen$^1$, E.E. Krasovskii$^{2,3,4}$,\\ Michiel J.A. de Dood$^1$, Rudolf M. Tromp$^{5,1}$,
        Milan P. Allan$^1$,\\ Sense Jan van der Molen$^{1,*}$} 
\date{%
\small{
$^1$Leiden Institute of Physics, Leiden University, P.O. Box 9504, 2300 RA Leiden, The Netherlands\\
$^2$Departamento de Pol\'imeros y Materiales Avanzados: F\'isica, 
Qu\'imica y Tecnolog\'ia, Universidad del Pais Vasco UPV/EHU, 20080 San Sebasti\'an/Donostia, Spain\\
$^3$IKERBASQUE, Basque Foundation for Science, E-48013 Bilbao, Spain\\
$^4$Donostia International Physics Center (DIPC), E-20018 San Sebasti\'an, Spain\\
$^5$IBM T.J.Watson Research Center, 1101 Kitchawan Road, P.O.\ Box 218, Yorktown Heights, New York, New York 10598, USA}
\today}

\begin{document}

\maketitle

\section{Introduction}

{\bf%
In twisted bilayer graphene (TBG) a moir\'e pattern forms that introduces a new length scale to the material. At the 'magic' twist angle $\theta_m \approx 1.1^\circ$, this causes a flat band to form, yielding emergent properties such as correlated insulator behavior and superconductivity~\cite{Cao2018, Bistritzer2011, Lu2019efetov, Lisi2020}.
In general, the moiré structure in TBG varies spatially, 
influencing the local electronic properties~\cite{qiao_twisted_2018, khatibi_strain_2019, bi_designing_2019,Parker2021,Tilak2021} and hence the outcome of macroscopic charge transport experiments. 
In particular, to understand the wide variety observed in the phase diagrams and critical temperatures, a more detailed understanding of the local moir\'e variation is needed~\cite{AndreiMacDonald2020}.
Here, we study spatial and temporal variations of the moiré pattern in TBG using aberration-corrected Low Energy Electron Microscopy (AC-LEEM)~\cite{Tromp2010,Tromp2013}. 
The spatial variation we find is lower than reported previously. 
At 500\,$^\circ$C, we observe thermal fluctuations of the moiré lattice, corresponding to collective atomic displacements of less than 70\,pm on a time scale of seconds~\cite{Koshino2019}, homogenizing the sample.
Despite previous concerns, no untwisting of the layers is found, even at temperatures as high as 600\,$^\circ$C \cite{Choi2019,Cao2018correlated}. 
From these observations, we conclude that thermal annealing can be used to decrease the local disorder in TBG samples.
Finally, we report the existence of individual edge dislocations in the atomic and moir\'e lattice. 
These topological defects break translation symmetry and are anticipated to exhibit unique local electronic properties.  
\\
}

In charge transport experiments, a percolative average of the microscopic properties is measured. Therefore, any local variation in twist angle and strain in TBG will influence the result of such experiments.
However, imaging such microscopic variations is non-trivial. A myriad of experimental techniques has been applied to the problem~\cite{Xie2019Spectroscopic, Yoo2019, Uri2020SOT, halbertal2020moire, kazmierczak_strain_2021}, each only resolving part of the puzzle due to practical limitations (capping layer or device substrate, surface quality or measurement speed).

Here, we use (AC-)LEEM, which measures an image of the reflection of a micron-sized beam of electrons at a landing energy $E_0$ (0--100\,eV, referenced to the vacuum energy) in real space, in reciprocal space (diffraction), or combinations thereof. This allows us to perform large-scale, fast and non-destructive imaging of TBG.
Additionally, spectroscopic measurements, yielding information on the material's unoccupied bands can be done by varying $E_0$\cite{Flege2014,jobst2016quantifying}.

\section{Results}
\begin{figure}[!ht]
\centering
\includegraphics[width=\columnwidth]{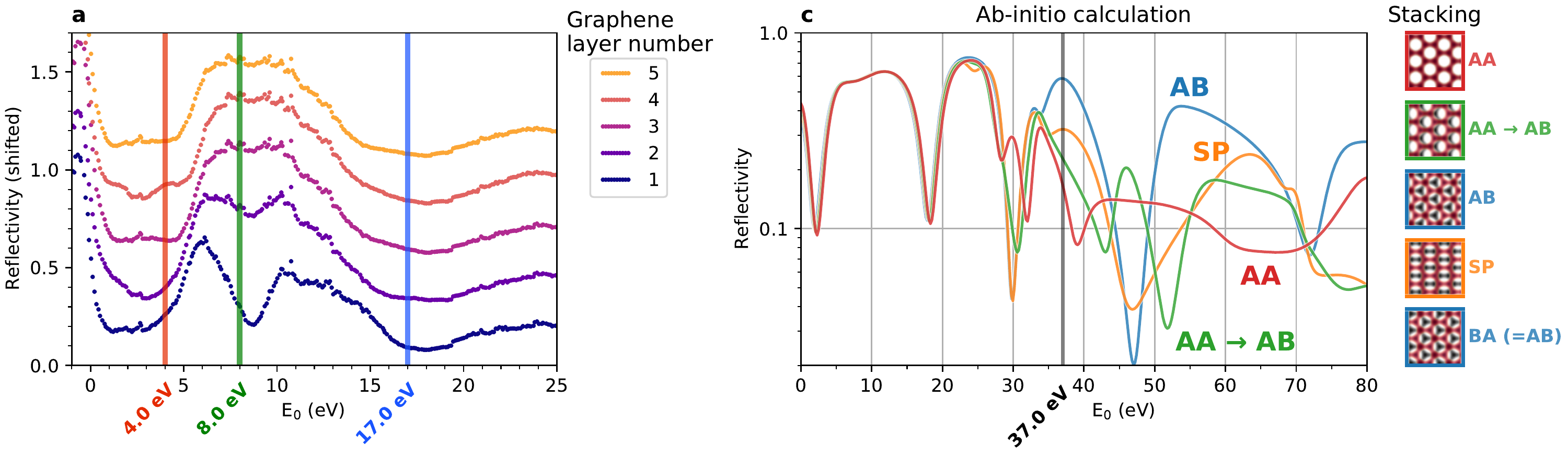}
\includegraphics[width=\columnwidth]{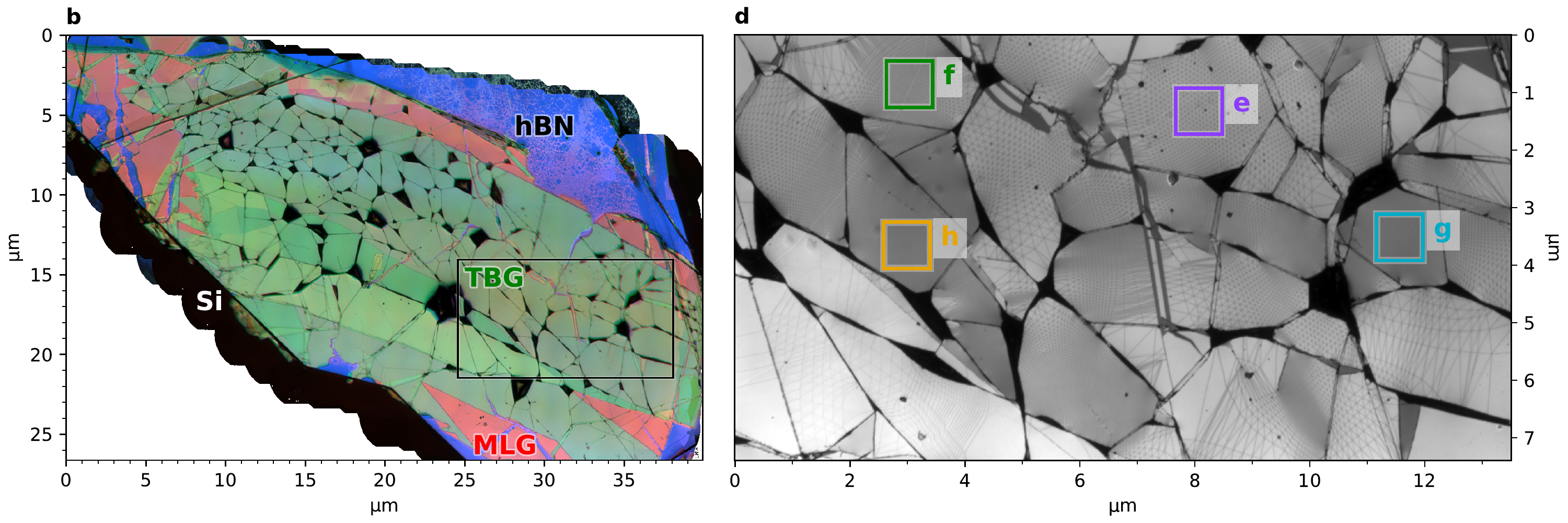}
\includegraphics[width=\columnwidth]{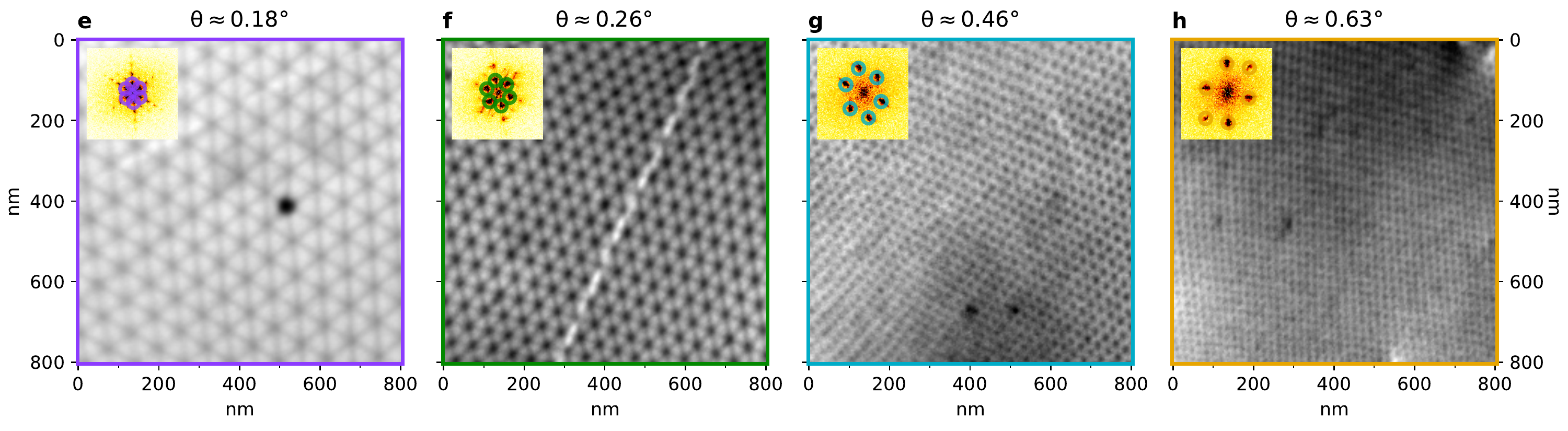}
\caption{\textbf{Device-scale imaging of TBG. }
\subf{a} Local spectra used to determine the graphene layer count.
\subf{b} Stitched composite bright field overview of a sample using 4\,eV, 8\,eV and 17\,eV as imaging energies in red, green and blue respectively (see main text for color interpretation). Visible defects include folds in black, tears, where the monolayer or even bare hBN shines through, bubbles (bright) and some polymer residue in the lower and upper right (dark speckles). 
\subf{c} Ab-initio calculations of LEEM spectra of different relative stackings of bilayer graphene, 37\,eV indicated.
\subf{d} Stitched bright field overview of the same sample imaged at $E_0=37\,\text{eV}$, for optimal stacking contrast, revealing the moir\'e patterns.
\subf{e-h} Crops of different twist angles areas from \subf{d} inset shows Fourier transforms and the detected moiré peaks. All measurements have been done on the same sample, featured in \textbf{b}.
}
\label{fig:overview}
\end{figure}

In \figref[a]{overview}, LEEM spectra are shown, taken on several locations of a TBG sample. 
These LEEM spectra are directly related to layer count, as described in refs.~\cite{Hibino2012Growth, feenstra_low-energy_2013, deJong2018measuring, jobst2016quantifying}; on the one hand via interlayer resonances in the 0--5\,eV range, on the other hand via the gradual disappearance of a minimum at 8\,eV. 
Here, more graphene layers (having a band gap at 8\,eV) are progressively masking an hBN band underneath. This allows us to determine the local graphene layer count for each point on our sample. To visualize that, we choose three characteristic energies, i.e. $E_0 = 4\eV$ (red), $E_0= 8\eV$ (green) and $E_0=17\eV$ (blue) (see  \figref[a]{overview}), and combine stitched overviews at these energies into a single false-color image (\figref[b]{overview}). This overview confirms that the sample consists of large TBG areas (darker green in \figref[b]{overview}) surrounded by monolayer graphene (pink), on an hBN flake (blue/purple) on silicon (black). 
Stripes of brighter green indicate areas of 2-on-2 graphene layers (lower stripe), 2-on-1  (upper stripe), and 1-on-2 (wedge on the lower right). The relatively homogeneous areas are themselves separated by folds, appearing as black lines. The folds locally combine in larger dark nodes (confirmed by AFM, see Supplementary Information \ref{sec:AFM}). 
A few folds, however, have folded over and appear as lines of higher layer count. Hence, \figref[b]{overview} provides a remarkable overview of a larger-scale sample, with detailed local information.\\
Increasing $E_0$ beyond $25\eV$, stacking boundaries and AA-sites become visible~\cite{Hibino2012Growth,dejong2018intrinsic}. 
This is consistent with ab-initio calculations of LEEM spectra for different relative stackings, as presented in \figref[c]{overview}. 
Therefore, imaging at $E_0=37\eV$ (indicated in \figref[c]{overview}) yields a precise map of the moiré lattice over the full TBG area  (see \figref[d]{overview}). We find that separate areas, between folds, exhibit different moiré periodicities and distortion~\cite{Beechem2014}. 
This allows us to study different moiré structures on a single sample. \figref[e-h]{overview} shows full resolution crops of areas indicated in \figref[d]{overview}. The observed twist angles on this sample range from $<0.1^\circ$ to 0.7$^\circ$.
For smaller angles, we observe local reconstruction towards Bernal stacking within the moir\'e lattice, consistent with literature~\cite{Yoo2019, kazmierczak_strain_2021}. 
The best resolution was reached on another sample with a twist angle of 1.3$^\circ$ (See Supplementary \figref{resolution}).

\subsection{Distortions \& Strain}

\begin{figure}[!ht]
\centering
\includegraphics[width=\columnwidth]{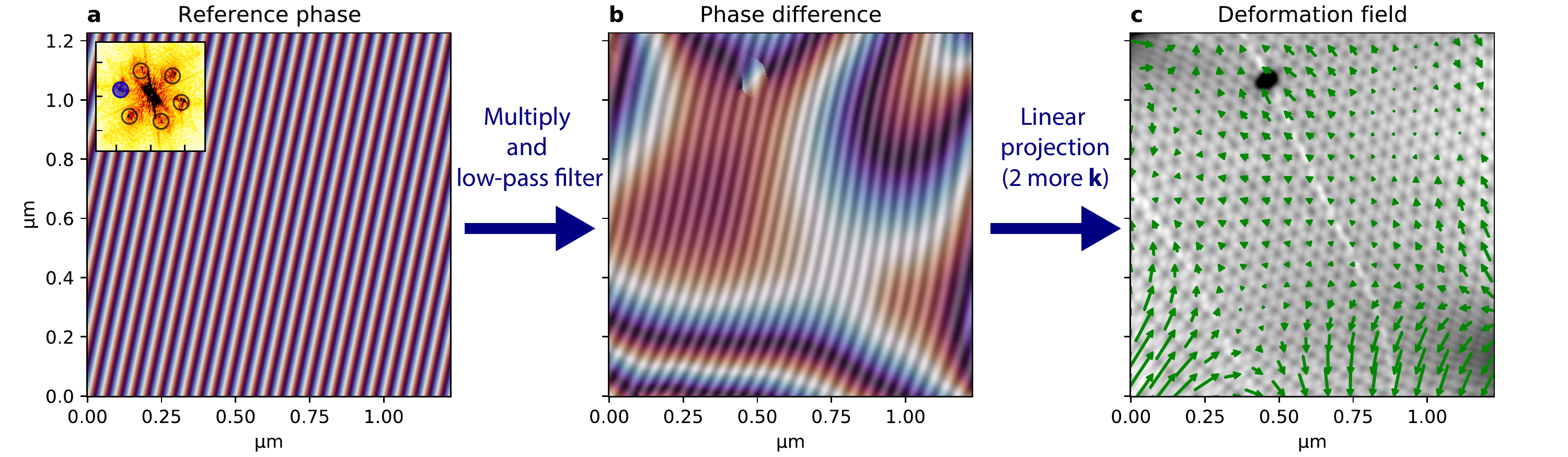}
\includegraphics[width=\columnwidth]{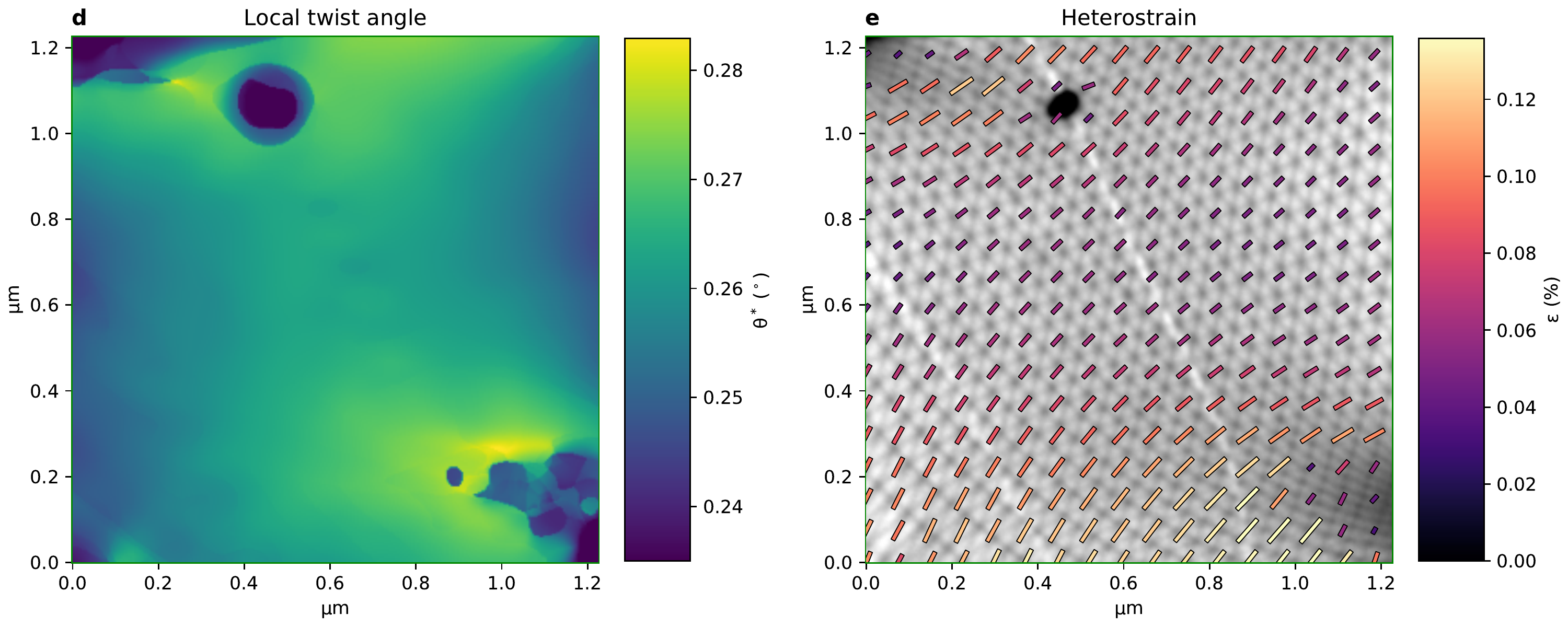}
\includegraphics[width=\columnwidth]{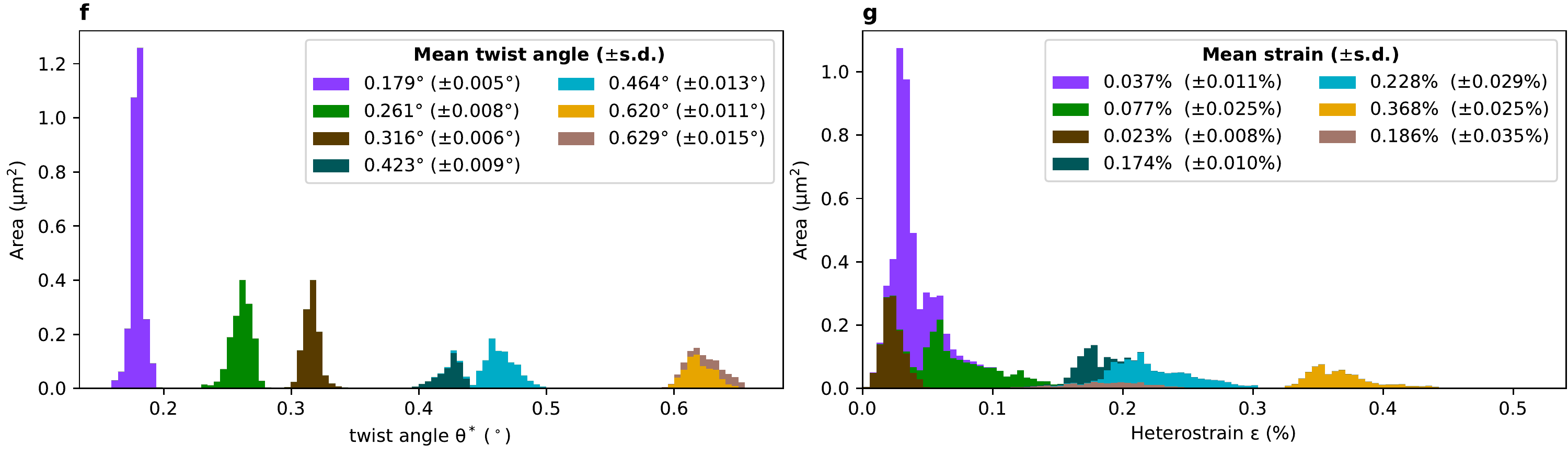}
\caption{\textbf{Distortion variation from Geometric Phase Analysis}
\subf{a} Reference phase corresponding to one $k$-vector used in GPA, as extracted from the Fourier transform of the corresponding image (inset).
\subf{b} Phase difference or GPA phase, obtained by multiplying with the original image and low-pass filtering. Overlayed is the corresponding extracted image wave.
\subf{c} The displacement field extracted from the GPA phases.
\subf{d} Extracted local twist angle $\theta^{*}$.
\subf{e} Extracted local heterostrain. Length and color of bars indicates the magnitude $\epsilon$ of the heterostrain, direction is the direction of elongation of the atomic lattice.
\subf{f,g} Distribution of $\theta^{*}$ \& $\epsilon$ extracted from different areas on the sample. Bar colors correspond to colors in \figref[d]{overview}, with the remaining areas shown in the supplementary information.
 }
\label{fig:distortion}
\end{figure}

The moir\'e patterns shows distortions, corresponding to local variations in twist angle and (interlayer) strain. Near folds, for instance, the strain increases resulting in strongly elongated triangles, for example in the lower right corner of \figref[d]{overview}~\cite{mao_evidence_2020}.
Despite their relative homogeneity, the moiré areas in \figref[e-h]{overview} also show subtle distortions.
 As structural variations correlate directly with local electronic properties, we quantify them in detail~\cite{qiao_twisted_2018, khatibi_strain_2019}.
For this, we use adaptive geometric phase analysis (GPA), extending our earlier work on STM data of moir\'e patterns in TBG (see Supplementary Information)~\cite{Hytch1998,Lawler2010,benschop2020mapping,Kemao2007,pyGPA}.
This method, illustrated in \figref[a-c]{distortion}, yields the displacement field with respect to a perfect lattice, by multiplying the original image with complex reference waves followed by low-pass filtering to obtain the GPA phase differences, which are then converted to the displacement field.
This field fully describes the distortion of the moir\'e lattice and allows us to extract key parameters such as the local twist angle $\theta^*(\vec r)$ (see \figref[d]{distortion}), and heterostrain magnitude $\epsilon(\vec r)$ and direction (see \figref[e]{distortion}).~\cite{Kerelsky2019,benschop2020mapping}
The distortions of the moiré pattern correspond directly to distortions of the atomic lattices, magnified by a factor $1/\theta$ and rotated by $90^\circ+\theta/2$~\cite{cosma_moire_2014,benschop2020mapping}. 

The extracted variation in twist angle and heterostrain for various regions of the sample, including those in \figref[e-h]{overview}, is summarized in \figref[f,g]{distortion}, respectively. The twist angle variation within each domain is much smaller than the variation in twist angle between the separate, fold-bounded areas. Within domains, standard deviations range from 0.005$^\circ$ to 0.015$^\circ$, i.e. significantly smaller (by a factor 3--10) than previously reported~\cite{Uri2020SOT, benschop2020mapping, kazmierczak_strain_2021}. 
The strain observed is around a few tenths of a percent, which is considerable. In some domains, we find an average strain of the atomic lattice of up to $\epsilon=0.4\,\%$. According to earlier theoretical work, such values are high enough to locally induce a quantum phase transition~\cite{Parker2021}.

The variation of $\epsilon$, as for $\theta^{*}$, within the domains is significantly lower than in earlier studies. 
We do note that the use of GPA introduces a point spread function (PSF) that is broader than the PSF of the instrument, resulting in a lower displacement field frequency response at small scales and therefore a somewhat reduced variation. Nevertheless, the combined PSF of instrument and analysis is still comparable to other techniques that do not image the unit cell directly, allowing for a direct comparison.

We hypothesize that the difference in variations with literature stems from the relatively high temperature at which we annealed and measured the sample, combined with the relatively long averaging time of this measurement ($\geq 16\,$s for all data in \figref{distortion}). The high temperature induces thermal fluctuations of the lattice (as demonstrated below), allowing the system to approach a lower energy, more homogeneous, state.

\subsection{Edge dislocations}
\begin{figure}[!ht]
\centering
\includegraphics[width=\columnwidth]{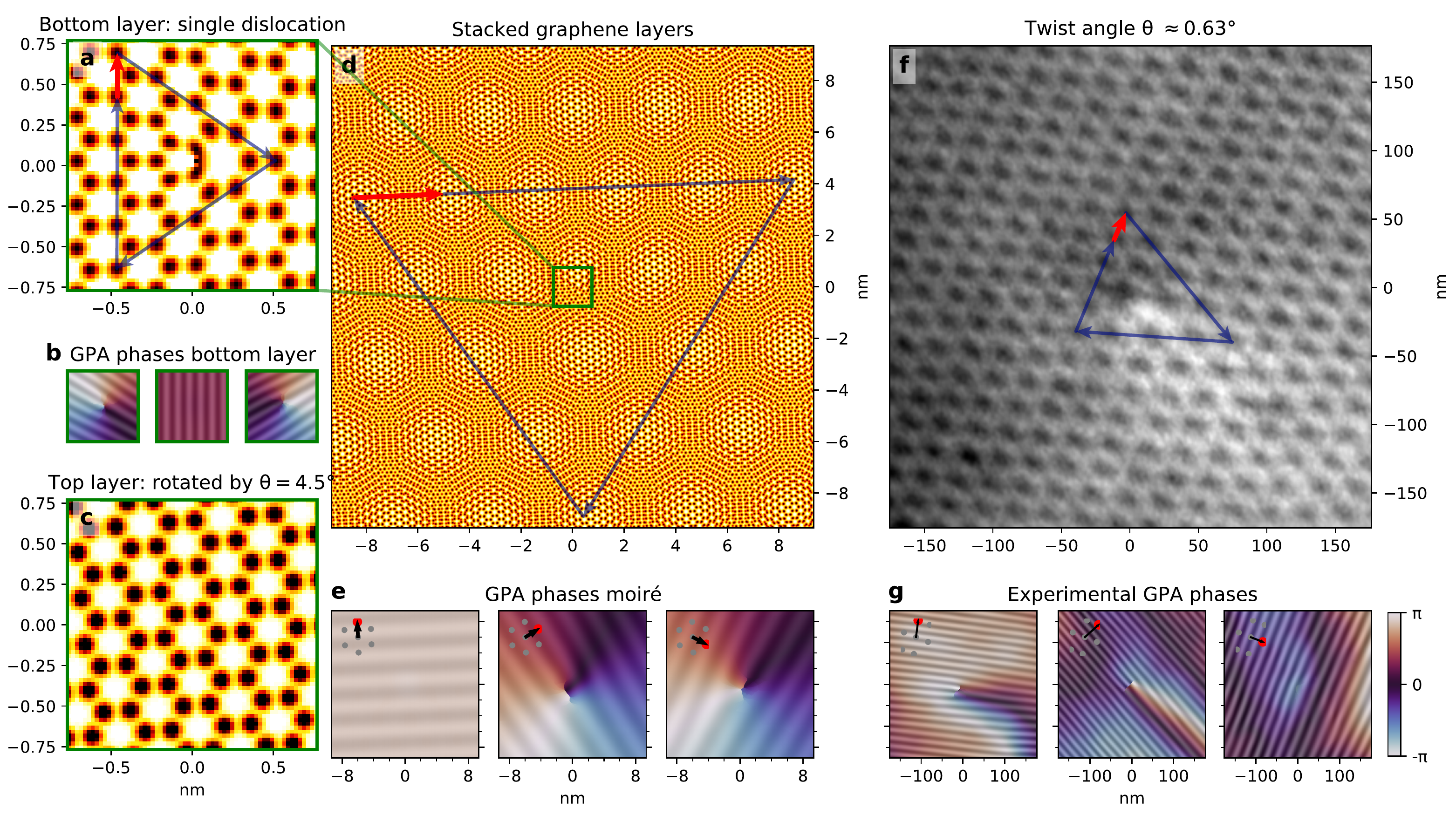}
\caption{\textbf{Edge dislocations in moir\'e systems}
\subf{a} Schematic of an edge dislocation in a single layer graphene (centered in the field-of-view), with the corresponding Burgers vector indicated in red.
\subf{b} GPA phases of \subref{a}.
\subf{c} Top layer without dislocation, rotated $4.5^\circ$ with respect to the layer in \subref{a}.
\subf{d} Schematic of an edge dislocation of a single layer graphene in a twisted bilayer system. The green square in the center corresponds to the combination of \textbf{a} and \textbf{c}. The Burgers vector of the moir\'e lattice defect is indicated in red.
\subf{e} GPA phases of \subref{d}, exhibiting a singularity in the center. The used moiré reference vectors are indicated.
\subf{f} LEEM image of an edge dislocation in a TBG moiré lattice with a twist angle $\theta \approx 0.63^\circ$. The corresponding moir\'e Burgers vector is indicated in red.
\subf{g} GPA phases corresponding to \subref{f}.
}
\label{fig:dislocation}
\end{figure}

So far, we have discussed structural properties varying on the moir\'e length scale. However, the moir\'e magnification of deformations is general and extends to atomic edge dislocations (visualized in \figref[a]{dislocation}). This type of topological defect stems from a missing row of atomic unit cells and is characterized by an in-plane Burgers vector (in red)~\cite{burgers1940geometrical,Lardner1971}. 
The addition of a second (twisted) atomic layer magnifies (and rotates) the defect to an edge dislocation in the moir\'e lattice (illustrated in \figref[d,e]{dislocation})~\cite{cosma_moire_2014}.
In all cases, the defect can be characterized using GPA, pinpointing its location and Burgers vector (\figref[b,e]{dislocation})~\cite{Warner2012}.

In our sample, we find a few such defects in the moir\'e lattice (see SI).
In \figref[f,g]{dislocation} we show an edge dislocation in a topographically flat region with $\theta=0.63^\circ$ (see SI). 
Contrary to TEM observations on single-layer graphene~\cite{Warner2012}, we do not observe creation or annihilation of edge dislocation pairs in the microscope, even at elevated measurement temperatures (500 $^\circ$C) and under low-energy electron irradiation. Furthermore, the mobility of the defects is low.
One edge dislocation did move, over several moir\'e cells between measurements, after which it remained at the same position even after a month at room temperature and reheating (see supplementary information).
This stability suggests that the moir\'e lattice itself plays a role in stabilizing these defects, via a minimum of the local stacking fault energy within the moir\'e unit cell. 

These topological dislocations break translational symmetry, which may lead to singular electronic properties on the local scale~\cite{Mesaros2009,liu_direct_2021,de_beule_aharonov-bohm_2020}. Specifically, a phase difference will appear between electron paths encircling the defect clockwise and counterclockwise.

\subsection{High temperature dynamics of the moiré lattice}

\begin{figure}[!ht]
\centering
\includegraphics[width=\columnwidth]{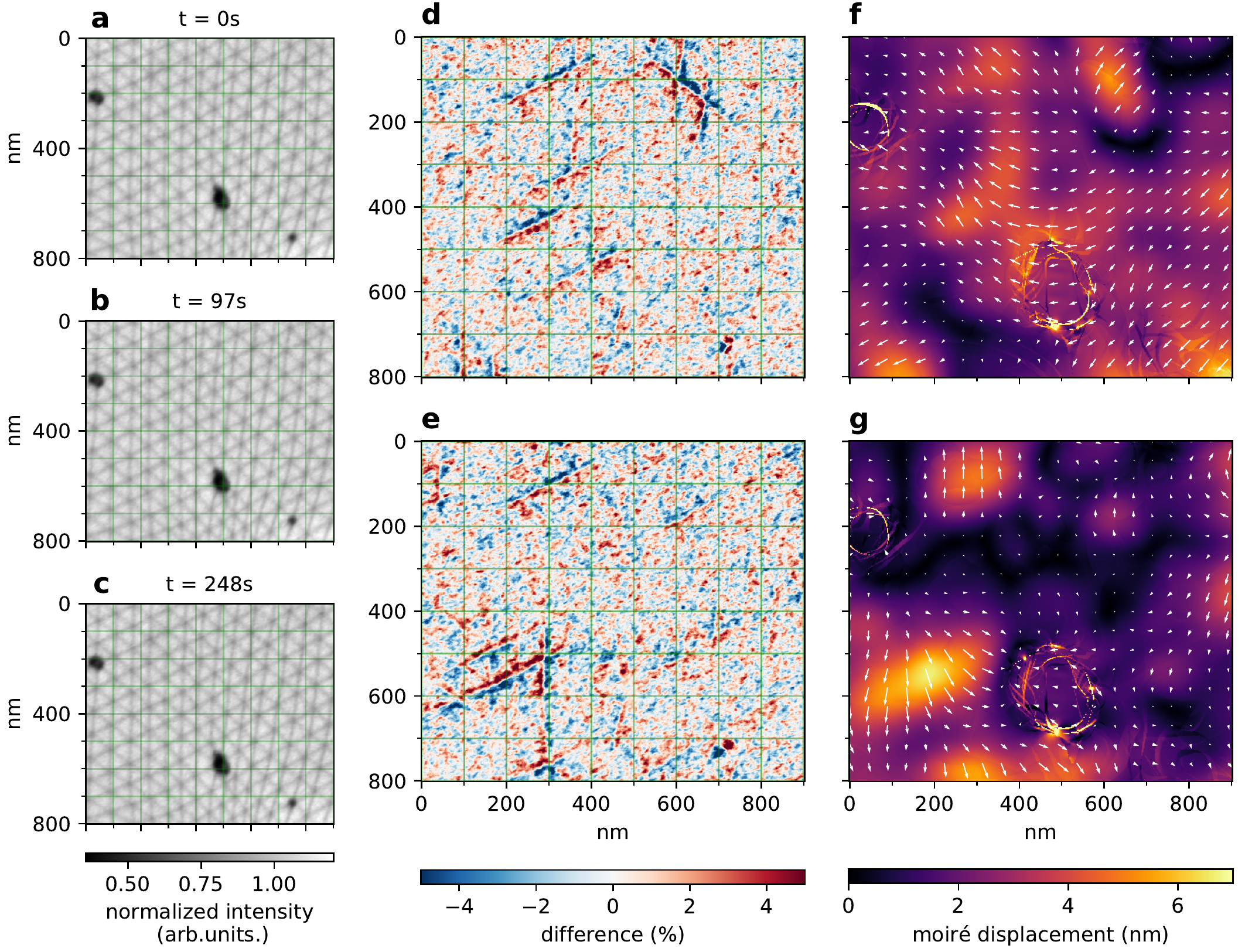}
\caption{\textbf{Dynamics of moir\'e patterns}
\subf{a-c} Three images of the same area, taken minutes apart at a constant temperature of 500\,$^\circ$C.
\subf{d,e} Difference of respectively \textbf{b} and \textbf{c} with \textbf{a}, i.e. $t=0\,\text{s}$, highlighting the shift of the domain boundaries.
\subf{f,g} GPA extracted displacement of respectively \subf{b} and \subf{c} with respect to $t=0\,\text{s}$, with the arrows indicating the direction and amplitude magnified 8 times for visibility.}
\label{fig:dynamics}
\end{figure}

All measurements presented so far were performed at 500\,$^\circ$C, to minimize hydrocarbon contamination under the electron beam. In literature, there is concern about the graphene layers untwisting at such temperatures, due to energy differences between different rotations~\cite{Choi2019,Cao2018correlated}. We, however, see no sign of that. The twist angles within the domains are stable from 100\,$^\circ$C up to 600\,$^\circ$C for all samples studied.
However, we did observe a more subtle thermal influence on the moiré pattern. At a temperature of 500\,$^\circ$C, the position of the stacking domain boundaries fluctuates slightly as a function of time (see \figref[a-c]{dynamics}).
Taking the difference of later images (\figref[b-c]{dynamics}) with the first image (\figref[a]{dynamics}), we clearly see the domain boundaries shifting (\figref[d,e]{dynamics}).
Moreover, we can quantify these fluctuations via the difference in displacement field with respect to the image at $t=0\,\text{s}$ using GPA (\figref[f,g]{dynamics}). Interestingly, these involve the collective movement of millions of atoms, but only over very small distances. 

We stress that a translation of the domain boundary by $4$\,nm, as observed, corresponds to a shift of less than half the width of the domain boundary itself~\cite{Yoo2019,gargiulo_structural_2017}. As the relative shift of the layers over the full domain boundary is a single carbon bond length, the corresponding atomic translations are less than half of that, i.e. less than $70$\,pm.
Hence, the `moiré magnification' makes it possible to detect these sub-Angstrom changes in TBG in real time using LEEM. Our data suggest that domain boundary displacement follows a random pattern of forward and backward steps. 
This indicates a possible source for the twist angle disorder observed in low(er) temperature experiments~\cite{kazmierczak_strain_2021,Uri2020SOT, benschop2020mapping,AndreiMacDonald2020}: frozen-in thermal fluctuations of the moir\'e lattice.
The thermal fluctuations found, corresponding to ${\pm0.005}^\circ$ for twist angle and $\pm 0.02 \%$ for strain, are smaller than the extracted static deformations, though not negligible. Note that these values are damped by the intrinsic broadening of GPA and the time integration.
Future experiments will focus on deducing the detailed statistics of the domain boundary dynamics versus temperature. 
Following these local collective excitations in time, will yield quantitative information on the energy landscape of these atomic lattice deformations within the moiré lattice. This will be important to answer the question if moiré lattices can be relaxed and homogenized using controlled annealing. If so, this would yield higher-quality magic-angle TBG devices in which charge transport is not limited by percolative effects and higher critical temperatures are reached.

\section{Conclusion}
Our quantitative LEEM study on TBG reveals a wide variety in twist angles and strain levels in a single sample.
We show that spontaneous changes in global twist angle do not occur, even during high-temperature annealing, but that local collective fluctuations do take place. This suggests that high-temperature annealing causes relaxation of the local moir\'e lattice, reducing lattice disorder. Vice versa, this points to frozen-in thermal fluctuations as a possible source of the (significant) short-range twist angle disorder observed previously. Furthermore, this potentially offers insight into energetic aspects of the atomic lattice deformation within the moiré lattice.

We also report the observation of stable topological defects, i.e. edge dislocations, in the moir\'e lattice of two Van der Waals layers. Combining our methods with other techniques that can access the electronic structure, such as STS, nanoARPES, and even in-situ potentiometry~\cite{kautz2015}, will allow for a systematic study of the electronic properties around these defects.
Finally, the methods we describe here extend beyond TBG, to any type of twisted system. Therefore, our work introduces a new way of studying deformations of moiré patterns and of connecting these to the (local) electronic properties of this exciting class of materials.

\section{Methods}
\subsection{Sample fabrication}\label{sec:samplefab}
The twisted bilayer graphene sample was fabricated using the standard tear-and-stack method ~\cite{Kim2017Tunable,Cao2018}. The monolayer graphene was first exfoliated with scotch tape on to a SiO$_2$/Si substrate. A polycarbonate (PC)/polydimethylsiloxane (PDMS) stamp was used for the transfer process, where the PC covered only half of the PDMS surface. After the first half of the graphene flake was successfully torn and picked up, it was rotated by 1.0$^{\circ}$. 
The flake was then overlapped with the bottom half and used to pick it up. The stack was then stamped on a moderately thick ($\sim$140nm) hBN flake, priorly exfoliated with PDMS on to a silicon substrate, along with the PC layer. Part of the graphene flake is deliberately put in contact with the Silicon surface for electrical contact purposes, i.e. to absorb the beam current. 
The whole substrate is then left in chloroform for 3 hours to dissolve the PC. 
All flakes were exfoliated from crystals, commercially bought from HQ Graphene and the fabrication process was performed using the Manual 2D Heterostructure Transfer System sold by the same company.

\subsection{LEEM}
All LEEM measurements where performed in the ESCHER LEEM, based on the SPECS P90~\cite{Tromp2010,Tromp2013,escherleem}.
Samples were loaded into the ultrahigh-vacuum (base pressure better than $1.0 \times 10^{-9} \text{mbar}$) LEEM main chamber and slowly heated to 500$^\circ$C (as measured by pyrometer and confirmed by IR-camera) and left to anneal overnight to get rid of any (polymer) residue.
All measurements were conducted at elevated temperatures, 450$^\circ$C -- 500$^\circ$C, unless specified otherwise.
The sample was located on the substrate using photo-emission electron microscopy (PEEM) with an unfiltered mercury short-arc lamp, by comparing to optical microscopy images taken beforehand. Spectra were taken in high-dynamic-range mode and drift corrected and all images were corrected for detector artefacts, as described in ref.~\cite{dejong2019quantitative}. When needed to obtain a sufficient signal-to-noise ratio, multiple $250$ ms exposures were accumulated for each image, e.g. 8 exposures (2 seconds) per landing energy in the spectra in \figref{overview}{b} and 16 exposures (4 seconds) at each location for the overview at 37\,eV in \figref{overview}{c,e}.

\subsubsection{Time series}
To measure the dynamics as presented in \figref{dynamics}{}, a time series of accumulated $4\times 250\,\text{ms} = 1\,\text{s}$ exposure images was taken back-to-back. After regular detector correction and drift correction, each image was divided by a Gaussian smoothed version of itself ($\sigma = 50\,\text{pixels}$) to get rid of spatial and temporal fluctuations in electron illumination intensity. To further reduce noise, a Gaussian filter with a width of $\sigma = 1\,\text{image} \sim 1\,\text{s}$, was applied in the time direction before applying GPA.

\subsection{Stitching}
To enable high resolution, large field-of-view LEEM imaging, the LEEM sample stage~\cite{Ellis2013} was scanned in a rectangular pattern over the sample, taking an image at each position, leaving sufficient overlap ($2\,\mu$m steps at a $4.7\,\mu$m field-of-view).
To obtain meaningful deformation information from this, care needs to be taken to use a stitching algorithm that does not introduce additional deformation, i.e. as faithfully reproducing reality as the constituting images.

To achieve this, a custom stitching algorithm tailored towards such LEEM data, was developed, as described in the Supplementary information and in the implementation~\cite{LEEM-analysis}.

In addition, for the composite bright field in \figref{overview}{a}, minor rotation and magnifications differences due to objective lens focus differences were compensated for. This was done by registering the stitches for different energies using a log-polar transformation based method to obtain relative rotations and magnification.
Subsequently, areas where a color channel was missing, were imputed using a $k$-nearest neighbor lookup in a regularly sliced subset of the area with all color channels present.

\subsection{Image analysis}\label{sec:extendedGPA}
To quantify the large deviations in lattice shape due to the moir\'e magnification of small lattice distortions, we extended the GPA algorithm to use an adaptive grid of reference wave vectors, based on related to earlier work in laser fringe analysis~\cite{Kemao2007}. 

The spatial lock-in signal is calculated for a grid of wave vectors around a base reference vector, converting the GPA phase to reference the base reference vector every time. For each pixel, the spatial lock-in signal with the highest amplitude is selected as the final signal.
To avoid the problem of globally consistent phase unwrapping, the gradient of each GPA phase was directly converted to the displacement gradient tensor. More details of the used algorithm are given in the Supplementary information.

All image analysis code was written in Python, using \texttt{Numpy}~\cite{harris2020array}, \texttt{Scipy}~\cite{2020SciPy-NMeth}, \texttt{scikit-image}~\cite{vanderWalt2014} and \texttt{Dask}~\cite{dask}. The core algorithms will be made available as an open source Python package~\cite{pyGPA}.  Throughout the development of the algorithms and writing of the paper, \texttt{matplotlib}~\cite{Hunter2007,matplotlib3_3_2} was extensively used for plotting and figure creation.

\subsection{Reflectivity calculations}
The theoretical reflectivity spectra are obtained with the ab-initio Bloch-wave-based
scattering method described in ref. \cite{Krasovskii2004}. Details of the application
of this method to stand-alone two-dimensional films of finite thickness can be found
in ref. \cite{Krasovskii2021}. The underlying all-electron Kohn-Sham potential was
obtained with a full-potential linear augmented plane-wave method within the local
density approximation, as explained in ref. \cite{Krasovskii1999}. Inelastic scattering
is taken into account by an absorbing imaginary potential $-iV_i$, which is taken to be
spatially constant ($V_i = 0.5\eV$) over a finite slab (where the electron density is
non-negligible) and to be zero in the two semi-infinite vacuum half spaces. In addition,
a Gaussian broadening of 1\,eV is applied to account for experimental losses.

\section{Acknowledgements}
We thank Marcel Hesselberth and Douwe Scholma for their indispensable technical support. We thank D. K. Efetov and J. Aarts for getting us started on twisted bilayer graphene.
We would like to thank J. Jobst and D.N.L. Kok for scientific discussions and L. Visser and G. Stam in particular for their input on the stitching algorithm.
We thank Federica Galli for both scientific discussion as well as technical support.
This work was supported by the Dutch Research Council (NWO) as part of the Frontiers of Nanoscience programme.
It was also supported by the Spanish Ministry of Science, Innovation and Universities (Project 
No. PID2019-105488GB-I00).

\section{Author contributions}
TAdJ performed the LEEM measurements and wrote the LEEM data analysis code. XC fabricated the sample and performed AFM measurements. 
TB contributed to the analysis. 
EK performed the reflectivity calculations. 
TAdJ and SJvdM took the initiative for the writing of the manuscript, with contributions from several others. 
All authors contributed to the scientific discussion.

\printbibliography

@article{Hibino2012Growth,
  doi = {10.1088/0022-3727/45/15/154008},
  url = {https://doi.org/10.1088/0022-3727/45/15/154008},
  year = {2012},
  month = mar,
  publisher = {{IOP} Publishing},
  volume = {45},
  number = {15},
  pages = {154008},
  author = {H Hibino and S Tanabe and S Mizuno and H Kageshima},
  title = {Growth and electronic transport properties of epitaxial graphene on {SiC}},
  journal = {Journal of Physics D: Applied Physics}
}

@article{jobst2016quantifying,
  doi = {10.1038/ncomms13621},
  url = {https://doi.org/10.1038/ncomms13621},
  year = {2016},
  month = nov,
  publisher = {Springer Science and Business Media {LLC}},
  volume = {7},
  number = {1},
  author = {Johannes Jobst and Alexander J. H. van der Torren and Eugene E. Krasovskii and Jesse Balgley and Cory R. Dean and Rudolf M. Tromp and Sense Jan van der Molen},
  title = {Quantifying electronic band interactions in van der Waals materials using angle-resolved reflected-electron spectroscopy},
  journal = {Nature Communications}
}

@article{dejong2018intrinsic,
  doi = {10.1103/physrevmaterials.2.104005},
  url = {https://doi.org/10.1103/physrevmaterials.2.104005},
  year = {2018},
  month = oct,
  publisher = {American Physical Society ({APS})},
  volume = {2},
  number = {10},
  author = {T. A. de Jong and E. E. Krasovskii and C. Ott and R. M. Tromp and S. J. van der Molen and J. Jobst},
  title = {Intrinsic stacking domains in graphene on silicon carbide: A pathway for intercalation},
  journal = {Physical Review Materials}
}

@article{deJong2018measuring,
  doi = {10.1002/pssb.201800191},
  url = {https://doi.org/10.1002/pssb.201800191},
  year = {2018},
  month = aug,
  publisher = {Wiley},
  volume = {255},
  number = {12},
  pages = {1800191},
  author = {Tobias A. de Jong and Johannes Jobst and Hyobin Yoo and Eugene E. Krasovskii and Philip Kim and Sense Jan van der Molen},
  title = {Measuring the Local Twist Angle and Layer Arrangement in Van der Waals Heterostructures},
  journal = {physica status solidi (b)}
}

@article{dejong2019quantitative,
  doi = {10.1016/j.ultramic.2019.112913},
  url = {https://doi.org/10.1016/j.ultramic.2019.112913},
  year = {2019},
  month = nov,
  publisher = {Elsevier {BV}},
  pages = {112913},
  author = {T.A. de Jong and D.N.L. Kok and A.J.H. van der Torren and H. Schopmans and R.M. Tromp and S.J. van der Molen and J. Jobst},
  title = {Quantitative analysis of spectroscopic Low Energy Electron Microscopy data: High-dynamic range imaging,  drift correction and cluster analysis},
  journal = {Ultramicroscopy}
}

@article{Hytch1998,
  doi = {10.1016/s0304-3991(98)00035-7},
  url = {https://doi.org/10.1016/s0304-3991(98)00035-7},
  year = {1998},
  month = aug,
  publisher = {Elsevier {BV}},
  volume = {74},
  number = {3},
  pages = {131--146},
  author = {M.J. Hÿtch and E. Snoeck and R. Kilaas},
  title = {Quantitative measurement of displacement and strain fields from {HREM} micrographs},
  journal = {Ultramicroscopy}
}

@article{Kemao2007,
  doi = {10.1016/j.optlaseng.2005.10.012},
  url = {https://doi.org/10.1016/j.optlaseng.2005.10.012},
  year = {2007},
  month = feb,
  publisher = {Elsevier {BV}},
  volume = {45},
  number = {2},
  pages = {304--317},
  author = {Qian Kemao},
  title = {Two-dimensional windowed Fourier transform for fringe pattern analysis: Principles,  applications and implementations},
  journal = {Optics and Lasers in Engineering}
}

@article{Tromp2010,
  doi = {10.1016/j.ultramic.2010.03.005},
  url = {https://doi.org/10.1016/j.ultramic.2010.03.005},
  year = {2010},
  month = jun,
  publisher = {Elsevier {BV}},
  volume = {110},
  number = {7},
  pages = {852--861},
  author = {R.M. Tromp and J.B. Hannon and A.W. Ellis and W. Wan and A. Berghaus and O. Schaff},
  title = {A new aberration-corrected,  energy-filtered {LEEM}/{PEEM} instrument. I. Principles and design},
  journal = {Ultramicroscopy}
}

@article{Tromp2013,
  doi = {10.1016/j.ultramic.2012.07.016},
  url = {https://doi.org/10.1016/j.ultramic.2012.07.016},
  year = {2013},
  month = apr,
  publisher = {Elsevier {BV}},
  volume = {127},
  pages = {25--39},
  author = {R.M. Tromp and J.B. Hannon and W. Wan and A. Berghaus and O. Schaff},
  title = {A new aberration-corrected,  energy-filtered {LEEM}/{PEEM} instrument {II}. Operation and results},
  journal = {Ultramicroscopy}
}

@article{escherleem,
  doi = {10.1147/jrd.2011.2150691},
  url = {https://doi.org/10.1147/jrd.2011.2150691},
  year = {2011},
  month = jul,
  publisher = {{IBM}},
  volume = {55},
  number = {4},
  pages = {1:1--1:7},
  author = {S. M. Schramm and J. Kautz and A. Berghaus and O. Schaff and R. M. Tromp and S. J. van der Molen},
  title = {Low-energy electron microscopy and spectroscopy with {ESCHER}: Status and prospects},
  journal = {{IBM} Journal of Research and Development}
}

@article{feenstra_low-energy_2013,
	title = {Low-energy electron reflectivity from graphene},
	volume = {87},
	url = {https://link.aps.org/doi/10.1103/PhysRevB.87.041406},
	doi = {10.1103/PhysRevB.87.041406},
	number = {4},
	journal = {Physical Review B},
	author = {Feenstra, R. M. and Srivastava, N. and Gao, Qin and Widom, M. and Diaconescu, Bogdan and Ohta, Taisuke and Kellogg, G. L. and Robinson, J. T. and Vlassiouk, I. V.},
	month = jan,
	year = {2013},
	note = {Publisher: American Physical Society},
	pages = {041406},
}

@article{kautz2015,
	title = {Low-{Energy} {Electron} {Potentiometry}: {Contactless} {Imaging} of {Charge} {Transport} on the {Nanoscale}},
	volume = {5},
	copyright = {2015 The Author(s)},
	issn = {2045-2322},
	shorttitle = {Low-{Energy} {Electron} {Potentiometry}},
	url = {https://www.nature.com/articles/srep13604},
	doi = {10.1038/srep13604},
	language = {en},
	number = {1},
	journal = {Scientific Reports},
	author = {Kautz, J. and Jobst, J. and Sorger, C. and Tromp, R. M. and Weber, H. B. and van der Molen, S. J.},
	month = sep,
	year = {2015},
	note = {Number: 1
Publisher: Nature Publishing Group},
	pages = {13604},
}

@article{         harris2020array,
 title         = {Array programming with {NumPy}},
 author        = {Charles R. Harris and K. Jarrod Millman and St{\'{e}}fan J.
                 van der Walt and Ralf Gommers and Pauli Virtanen and David
                 Cournapeau and Eric Wieser and Julian Taylor and Sebastian
                 Berg and Nathaniel J. Smith and Robert Kern and Matti Picus
                 and Stephan Hoyer and Marten H. van Kerkwijk and Matthew
                 Brett and Allan Haldane and Jaime Fern{\'{a}}ndez del
                 R{\'{i}}o and Mark Wiebe and Pearu Peterson and Pierre
                 G{\'{e}}rard-Marchant and Kevin Sheppard and Tyler Reddy and
                 Warren Weckesser and Hameer Abbasi and Christoph Gohlke and
                 Travis E. Oliphant},
 year          = {2020},
 month         = sep,
 journal       = {Nature},
 volume        = {585},
 number        = {7825},
 pages         = {357--362},
 doi           = {10.1038/s41586-020-2649-2},
 publisher     = {Springer Science and Business Media {LLC}},
 url           = {https://doi.org/10.1038/s41586-020-2649-2}
}

@article{2020SciPy-NMeth,
  author  = {Virtanen, Pauli and Gommers, Ralf and Oliphant, Travis E. and
            Haberland, Matt and Reddy, Tyler and Cournapeau, David and
            Burovski, Evgeni and Peterson, Pearu and Weckesser, Warren and
            Bright, Jonathan and {van der Walt}, St{\'e}fan J. and
            Brett, Matthew and Wilson, Joshua and Millman, K. Jarrod and
            Mayorov, Nikolay and Nelson, Andrew R. J. and Jones, Eric and
            Kern, Robert and Larson, Eric and Carey, C J and
            Polat, {\.I}lhan and Feng, Yu and Moore, Eric W. and
            {VanderPlas}, Jake and Laxalde, Denis and Perktold, Josef and
            Cimrman, Robert and Henriksen, Ian and Quintero, E. A. and
            Harris, Charles R. and Archibald, Anne M. and
            Ribeiro, Ant{\^o}nio H. and Pedregosa, Fabian and
            {van Mulbregt}, Paul and {SciPy 1.0 Contributors}},
  title   = {{{SciPy} 1.0: Fundamental Algorithms for Scientific
            Computing in Python}},
  journal = {Nature Methods},
  year    = {2020},
  volume  = {17},
  pages   = {261--272},
  adsurl  = {https://rdcu.be/b08Wh},
  doi     = {10.1038/s41592-019-0686-2},
}

@article{vanderWalt2014,
  doi = {10.7717/peerj.453},
  url = {https://doi.org/10.7717/peerj.453},
  year = {2014},
  month = jun,
  publisher = {{PeerJ}},
  volume = {2},
  pages = {e453},
  author = {St{\'{e}}fan van der Walt and Johannes L. Sch\"{o}nberger and Juan Nunez-Iglesias and Fran{\c{c}}ois Boulogne and Joshua D. Warner and Neil Yager and Emmanuelle Gouillart and Tony Yu},
  title = {scikit-image: image processing in Python},
  journal = {{PeerJ}}
}

@Manual{dask,
  title = {Dask: Library for dynamic task scheduling},
  author = {{Dask Development Team}},
  year = {2016},
  url = {https://dask.org},
}

@article{Hunter2007,
  Author    = {Hunter, J. D.},
  Title     = {Matplotlib: A 2D graphics environment},
  Journal   = {Computing in Science \& Engineering},
  Volume    = {9},
  Number    = {3},
  Pages     = {90--95},
  publisher = {IEEE COMPUTER SOC},
  doi       = {10.1109/MCSE.2007.55},
  year      = 2007
}

@misc{matplotlib3_3_2,
  doi = {10.5281/ZENODO.4030140},
  url = {https://zenodo.org/record/4030140},
  author = {Caswell,  Thomas A and Droettboom,  Michael and Lee,  Antony and Hunter,  John and Andrade,  Elliott Sales De and Firing,  Eric and Hoffmann,  Tim and Klymak,  Jody and Stansby,  David and Varoquaux,  Nelle and Nielsen,  Jens Hedegaard and Root,  Benjamin and May,  Ryan and Elson,  Phil and Sepp\"{a}nen,  Jouni K. and Dale,  Darren and {Jae-Joon Lee} and McDougall,  Damon and Straw,  Andrew and Hobson,  Paul and Gohlke,  Christoph and Yu,  Tony S and Ma,  Eric and Vincent,  Adrien F. and Silvester,  Steven and Moad,  Charlie and Kniazev,  Nikita and {,  Hannah} and Ernest,  Elan and Ivanov,  Paul},
  title = {matplotlib/matplotlib: REL: v3.3.2},
  publisher = {Zenodo},
  year = {2020},
  copyright = {Open Access}
}

@article{Ellis2013,
  doi = {10.1063/1.4813739},
  url = {https://doi.org/10.1063/1.4813739},
  year = {2013},
  month = jul,
  publisher = {{AIP} Publishing},
  volume = {84},
  number = {7},
  pages = {075112},
  author = {A. W. Ellis and R. M. Tromp},
  title = {A versatile ultra high vacuum sample stage with six degrees of freedom},
  journal = {Review of Scientific Instruments}
}

@article{Krasovskii2004,
author = {Krasovskii, E. E.},
doi = {10.1103/PhysRevB.70.245322},
journal = {Physical Review B},
pages = {245322},
title = {{Augmented-plane-wave approach to scattering of Bloch 
electrons by an interface}},
volume = {70},
year = {2004}
}

@article{Krasovskii1999,
author = {Krasovskii, E. E. and Starrost, F. and Schattke, W.},
doi = {10.1103/PhysRevB.59.10504},
journal = {Physical Review B},
pages = {10504},
title = {{Augmented Fourier components method for constructing the 
crystal potential in self-consistent band-structure calculations}},
volume = {59},
year = {1999}
}

@article{Krasovskii2021,
AUTHOR = {Krasovskii, Eugene},
TITLE = {Ab Initio Theory of Photoemission from Graphene},
JOURNAL = {Nanomaterials},
VOLUME = {11},
YEAR = {2021},
NUMBER = {5},
ARTICLE-NUMBER = {1212},
URL = {https://www.mdpi.com/2079-4991/11/5/1212},
ISSN = {2079-4991},
DOI = {10.3390/nano11051212}
}

@misc{pyGPA,
  author = {Tobias A. de Jong},
  title = {pyGPA},
  year  = {2021},
  url   = {https://github.com/TAdeJong/pyGPA},
}

@inproceedings{numba,
  doi = {10.1145/2833157.2833162},
  url = {https://doi.org/10.1145/2833157.2833162},
  year = {2015},
  publisher = {{ACM} Press},
  author = {Siu Kwan Lam and Antoine Pitrou and Stanley Seibert},
  title = {Numba},
  booktitle = {Proceedings of the Second Workshop on the {LLVM} Compiler Infrastructure in {HPC} - {LLVM} {\textquotesingle}15}
}

@misc{LEEM-analysis,
  doi = {10.5281/ZENODO.3539538},
  url = {https://zenodo.org/record/3539538},
  author = {Jong,  Tobias A. De},
  title = {TAdeJong/LEEM-analysis: LEEM-analysis},
  publisher = {Zenodo},
  year = {2019},
  copyright = {MIT License}
}

@article{Flege2014,
  doi = {10.1002/pssr.201409102},
  url = {https://doi.org/10.1002/pssr.201409102},
  year = {2014},
  month = may,
  publisher = {Wiley},
  volume = {8},
  number = {6},
  pages = {463--477},
  author = {Jan Ingo Flege and Eugene E. Krasovskii},
  title = {Intensity-voltage low-energy electron microscopy for functional materials characterization},
  journal = {physica status solidi ({RRL}) - Rapid Research Letters}
}

@article{AndreiMacDonald2020,
  doi = {10.1038/s41563-020-00840-0},
  url = {https://doi.org/10.1038/s41563-020-00840-0},
  year = {2020},
  month = nov,
  publisher = {Springer Science and Business Media {LLC}},
  volume = {19},
  number = {12},
  pages = {1265--1275},
  author = {Eva Y. Andrei and Allan H. MacDonald},
  title = {Graphene bilayers with a twist},
  journal = {Nature Materials}
}

@article{halbertal2020moire,
  doi = {10.1038/s41467-020-20428-1},
  url = {https://doi.org/10.1038/s41467-020-20428-1},
  year = {2021},
  month = jan,
  publisher = {Springer Science and Business Media {LLC}},
  volume = {12},
  number = {1},
  author = {Dorri Halbertal and Nathan R. Finney and Sai S. Sunku and Alexander Kerelsky and Carmen Rubio-Verd{\'{u}} and Sara Shabani and Lede Xian and Stephen Carr and Shaowen Chen and Charles Zhang and Lei Wang and Derick Gonzalez-Acevedo and Alexander S. McLeod and Daniel Rhodes and Kenji Watanabe and Takashi Taniguchi and Efthimios Kaxiras and Cory R. Dean and James C. Hone and Abhay N. Pasupathy and Dante M. Kennes and Angel Rubio and D. N. Basov},
  title = {Moir{\'{e}} metrology of energy landscapes in van der Waals heterostructures},
  journal = {Nature Communications}
}

@article{Yoo2019,
  doi = {10.1038/s41563-019-0346-z},
  url = {https://doi.org/10.1038/s41563-019-0346-z},
  year = {2019},
  month = apr,
  publisher = {Springer Science and Business Media {LLC}},
  volume = {18},
  number = {5},
  pages = {448--453},
  author = {Hyobin Yoo and Rebecca Engelke and Stephen Carr and Shiang Fang and Kuan Zhang and Paul Cazeaux and Suk Hyun Sung and Robert Hovden and Adam W. Tsen and Takashi Taniguchi and Kenji Watanabe and Gyu-Chul Yi and Miyoung Kim and Mitchell Luskin and Ellad B. Tadmor and Efthimios Kaxiras and Philip Kim},
  title = {Atomic and electronic reconstruction at the van der Waals interface in twisted bilayer graphene},
  journal = {Nature Materials}
}

@article{benschop2020mapping,
  doi = {10.1103/physrevresearch.3.013153},
  url = {https://doi.org/10.1103/physrevresearch.3.013153},
  year = {2021},
  month = feb,
  publisher = {American Physical Society ({APS})},
  volume = {3},
  number = {1},
  author = {Tjerk Benschop and Tobias A. de Jong and Petr Stepanov and Xiaobo Lu and Vincent Stalman and Sense Jan van der Molen and Dmitri K. Efetov and Milan P. Allan},
  title = {Measuring local moir{\'{e}} lattice heterogeneity of twisted bilayer graphene},
  journal = {Physical Review Research}
}

@article{Mesaros2009,
  doi = {10.1103/physrevb.79.155111},
  url = {https://doi.org/10.1103/physrevb.79.155111},
  year = {2009},
  month = apr,
  publisher = {American Physical Society ({APS})},
  volume = {79},
  number = {15},
  author = {A. Mesaros and D. Sadri and J. Zaanen},
  title = {Berry phase of dislocations in graphene and valley conserving decoherence},
  journal = {Physical Review B}
}

@article{Warner2012,
  doi = {10.1126/science.1217529},
  url = {https://doi.org/10.1126/science.1217529},
  year = {2012},
  month = jul,
  publisher = {American Association for the Advancement of Science ({AAAS})},
  volume = {337},
  number = {6091},
  pages = {209--212},
  author = {J. H. Warner and E. R. Margine and M. Mukai and A. W. Robertson and F. Giustino and A. I. Kirkland},
  title = {Dislocation-Driven Deformations in Graphene},
  journal = {Science}
}

@article{Bistritzer2011,
  doi = {10.1073/pnas.1108174108},
  url = {https://doi.org/10.1073/pnas.1108174108},
  year = {2011},
  month = jul,
  publisher = {Proceedings of the National Academy of Sciences},
  volume = {108},
  number = {30},
  pages = {12233--12237},
  author = {R. Bistritzer and A. H. MacDonald},
  title = {Moire bands in twisted double-layer graphene},
  journal = {Proceedings of the National Academy of Sciences}
}

@article{Uri2020SOT,
  doi = {10.1038/s41586-020-2255-3},
  url = {https://doi.org/10.1038/s41586-020-2255-3},
  year = {2020},
  month = may,
  publisher = {Springer Science and Business Media {LLC}},
  volume = {581},
  number = {7806},
  pages = {47--52},
  author = {A. Uri and S. Grover and Y. Cao and J.~A. Crosse and K. Bagani and D. Rodan-Legrain and Y. Myasoedov and K. Watanabe and T. Taniguchi and P. Moon and M. Koshino and P. Jarillo-Herrero and E. Zeldov},
  title = {Mapping the twist-angle disorder and Landau levels in magic-angle graphene},
  journal = {Nature}
}

@article{Xie2019Spectroscopic,
  doi = {10.1038/s41586-019-1422-x},
  url = {https://doi.org/10.1038/s41586-019-1422-x},
  year = {2019},
  month = jul,
  publisher = {Springer Science and Business Media {LLC}},
  volume = {572},
  number = {7767},
  pages = {101--105},
  author = {Yonglong Xie and Biao Lian and Berthold J\"{a}ck and Xiaomeng Liu and Cheng-Li Chiu and Kenji Watanabe and Takashi Taniguchi and B. Andrei Bernevig and Ali Yazdani},
  title = {Spectroscopic signatures of many-body correlations in magic-angle twisted bilayer graphene},
  journal = {Nature}
}

@article{Lu2019efetov,
  doi = {10.1038/s41586-019-1695-0},
  url = {https://doi.org/10.1038/s41586-019-1695-0},
  year = {2019},
  month = oct,
  publisher = {Springer Science and Business Media {LLC}},
  volume = {574},
  number = {7780},
  pages = {653--657},
  author = {Xiaobo Lu and Petr Stepanov and Wei Yang and Ming Xie and Mohammed Ali Aamir and Ipsita Das and Carles Urgell and Kenji Watanabe and Takashi Taniguchi and Guangyu Zhang and Adrian Bachtold and Allan H. MacDonald and Dmitri K. Efetov},
  title = {Superconductors,  orbital magnets and correlated states in magic-angle bilayer graphene},
  journal = {Nature}
}

@article{Cao2018,
  doi = {10.1038/nature26160},
  url = {https://doi.org/10.1038/nature26160},
  year = {2018},
  month = mar,
  publisher = {Springer Science and Business Media {LLC}},
  volume = {556},
  number = {7699},
  pages = {43--50},
  author = {Yuan Cao and Valla Fatemi and Shiang Fang and Kenji Watanabe and Takashi Taniguchi and Efthimios Kaxiras and Pablo Jarillo-Herrero},
  title = {Unconventional superconductivity in magic-angle graphene superlattices},
  journal = {Nature}
}

@article{Cao2018correlated,
  doi = {10.1038/nature26154},
  url = {https://doi.org/10.1038/nature26154},
  year = {2018},
  month = mar,
  publisher = {Springer Science and Business Media {LLC}},
  volume = {556},
  number = {7699},
  pages = {80--84},
  author = {Yuan Cao and Valla Fatemi and Ahmet Demir and Shiang Fang and Spencer L. Tomarken and Jason Y. Luo and Javier D. Sanchez-Yamagishi and Kenji Watanabe and Takashi Taniguchi and Efthimios Kaxiras and Ray C. Ashoori and Pablo Jarillo-Herrero},
  title = {Correlated insulator behaviour at half-filling in magic-angle graphene superlattices},
  journal = {Nature}
}

@article{liu_direct_2021,
	title = {Direct observation of magneto-electric {Aharonov}-{Bohm} effect in moir{\textbackslash}'e-scale quantum paths of minimally twisted bilayer graphene},
	url = {http://arxiv.org/abs/2102.00164},
	journal = {arXiv:2102.00164 [cond-mat]},
	author = {Liu, Yi-Wen and Ren, Ya-Ning and Hao, Chen-Yue and He, Lin},
	month = jan,
	year = {2021},
	note = {arXiv: 2102.00164},
}

@article{de_beule_aharonov-bohm_2020,
	title = {Aharonov-{Bohm} {Oscillations} in {Minimally} {Twisted} {Bilayer} {Graphene}},
	volume = {125},
	url = {https://link.aps.org/doi/10.1103/PhysRevLett.125.096402},
	doi = {10.1103/PhysRevLett.125.096402},
	number = {9},
	journal = {Physical Review Letters},
	author = {De Beule, C. and Dominguez, F. and Recher, P.},
	month = aug,
	year = {2020},
	note = {Publisher: American Physical Society},
	pages = {096402},
}

@article{kazmierczak_strain_2021,
	title = {Strain fields in twisted bilayer graphene},
	copyright = {2021 The Author(s), under exclusive licence to Springer Nature Limited},
	issn = {1476-4660},
	url = {https://www.nature.com/articles/s41563-021-00973-w},
	doi = {10.1038/s41563-021-00973-w},
	language = {en},
	journal = {Nature Materials},
	author = {Kazmierczak, Nathanael P. and Van Winkle, Madeline and Ophus, Colin and Bustillo, Karen C. and Carr, Stephen and Brown, Hamish G. and Ciston, Jim and Taniguchi, Takashi and Watanabe, Kenji and Bediako, D. Kwabena},
	month = apr,
	year = {2021},
	note = {Publisher: Nature Publishing Group},
	pages = {1--8},
}

@article{gargiulo_structural_2017,
	title = {Structural and electronic transformation in low-angle twisted bilayer graphene},
	volume = {5},
	issn = {2053-1583},
	url = {https://doi.org/10.1088/2053-1583/aa9640},
	doi = {10.1088/2053-1583/aa9640},
	language = {en},
	number = {1},
	journal = {2D Materials},
	author = {Gargiulo, Fernando and Yazyev, Oleg V.},
	month = nov,
	year = {2017},
	note = {Publisher: IOP Publishing},
	pages = {015019},
}

@article{Lawler2010,
  doi = {10.1038/nature09169},
  url = {https://doi.org/10.1038/nature09169},
  year = {2010},
  month = jul,
  publisher = {Springer Science and Business Media {LLC}},
  volume = {466},
  number = {7304},
  pages = {347--351},
  author = {M. J. Lawler and K. Fujita and Jhinhwan Lee and A. R. Schmidt and Y. Kohsaka and Chung Koo Kim and H. Eisaki and S. Uchida and J. C. Davis and J. P. Sethna and Eun-Ah Kim},
  title = {Intra-unit-cell electronic nematicity of the high-Tc copper-oxide pseudogap states},
  journal = {Nature}
}

@article{Lisi2020,
  doi = {10.1038/s41567-020-01041-x},
  url = {https://doi.org/10.1038/s41567-020-01041-x},
  year = {2020},
  month = sep,
  publisher = {Springer Science and Business Media {LLC}},
  volume = {17},
  number = {2},
  pages = {189--193},
  author = {Simone Lisi and Xiaobo Lu and Tjerk Benschop and Tobias A. de Jong and Petr Stepanov and Jose R. Duran and Florian Margot and Ir{\`{e}}ne Cucchi and Edoardo Cappelli and Andrew Hunter and Anna Tamai and Viktor Kandyba and Alessio Giampietri and Alexei Barinov and Johannes Jobst and Vincent Stalman and Maarten Leeuwenhoek and Kenji Watanabe and Takashi Taniguchi and Louk Rademaker and Sense Jan van der Molen and Milan P. Allan and Dmitri K. Efetov and Felix Baumberger},
  title = {Observation of flat bands in twisted bilayer graphene},
  journal = {Nature Physics}
}

@article{Choi2019,
  doi = {10.1038/s41567-019-0606-5},
  url = {https://doi.org/10.1038/s41567-019-0606-5},
  year = {2019},
  month = aug,
  publisher = {Springer Science and Business Media {LLC}},
  volume = {15},
  number = {11},
  pages = {1174--1180},
  author = {Youngjoon Choi and Jeannette Kemmer and Yang Peng and Alex Thomson and Harpreet Arora and Robert Polski and Yiran Zhang and Hechen Ren and Jason Alicea and Gil Refael and Felix von Oppen and Kenji Watanabe and Takashi Taniguchi and Stevan Nadj-Perge},
  title = {Electronic correlations in twisted bilayer graphene near the magic angle},
  journal = {Nature Physics}
}

@article{Kim2017Tunable,
  doi = {10.1073/pnas.1620140114},
  url = {https://doi.org/10.1073/pnas.1620140114},
  year = {2017},
  month = mar,
  publisher = {Proceedings of the National Academy of Sciences},
  volume = {114},
  number = {13},
  pages = {3364--3369},
  author = {Kyounghwan Kim and Ashley DaSilva and Shengqiang Huang and Babak Fallahazad and Stefano Larentis and Takashi Taniguchi and Kenji Watanabe and Brian J. LeRoy and Allan H. MacDonald and Emanuel Tutuc},
  title = {Tunable moir{\'{e}} bands and strong correlations in small-twist-angle bilayer graphene},
  journal = {Proceedings of the National Academy of Sciences}
}

@article{cosma_moire_2014,
	title = {Moiré pattern as a magnifying glass for strain and dislocations in van der {Waals} heterostructures},
	volume = {173},
	issn = {1364-5498},
	url = {https://pubs.rsc.org/en/content/articlelanding/2014/fd/c4fd00146j},
	doi = {10.1039/C4FD00146J},
	language = {en},
	number = {0},
	journal = {Faraday Discussions},
	author = {Cosma, Diana A. and Wallbank, John R. and Cheianov, Vadim and Fal'ko, Vladimir I.},
	month = dec,
	year = {2014},
	note = {Publisher: The Royal Society of Chemistry},
	pages = {137--143},
}

@article{qiao_twisted_2018,
	title = {Twisted graphene bilayer around the first magic angle engineered by heterostrain},
	volume = {98},
	url = {https://link.aps.org/doi/10.1103/PhysRevB.98.235402},
	doi = {10.1103/PhysRevB.98.235402},
	number = {23},
	journal = {Physical Review B},
	author = {Qiao, Jia-Bin and Yin, Long-Jing and He, Lin},
	month = dec,
	year = {2018},
	note = {Publisher: American Physical Society},
	pages = {235402},
}

@article{khatibi_strain_2019,
	title = {Strain impacts on commensurate bilayer graphene superlattices: {Distorted} trigonal warping, emergence of bandgap and direct-indirect bandgap transition},
	volume = {92},
	issn = {0925-9635},
	shorttitle = {Strain impacts on commensurate bilayer graphene superlattices},
	url = {https://www.sciencedirect.com/science/article/pii/S0925963518306241},
	doi = {10.1016/j.diamond.2018.12.007},
	language = {en},
	journal = {Diamond and Related Materials},
	author = {Khatibi, Zahra and Namiranian, Afshin and Parhizgar, Fariborz},
	month = feb,
	year = {2019},
	pages = {228--234},
}

@article{mao_evidence_2020,
	title = {Evidence of flat bands and correlated states in buckled graphene superlattices},
	volume = {584},
	copyright = {2020 The Author(s), under exclusive licence to Springer Nature Limited},
	issn = {1476-4687},
	url = {https://www.nature.com/articles/s41586-020-2567-3},
	doi = {10.1038/s41586-020-2567-3},
	language = {en},
	number = {7820},
	journal = {Nature},
	author = {Mao, Jinhai and Milovanović, Slaviša P. and Anđelković, Miša and Lai, Xinyuan and Cao, Yang and Watanabe, Kenji and Taniguchi, Takashi and Covaci, Lucian and Peeters, Francois M. and Geim, Andre K. and Jiang, Yuhang and Andrei, Eva Y.},
	month = aug,
	year = {2020},
	note = {Number: 7820
Publisher: Nature Publishing Group},
	pages = {215--220},
}

@article{bi_designing_2019,
	title = {Designing flat bands by strain},
	volume = {100},
	url = {https://link.aps.org/doi/10.1103/PhysRevB.100.035448},
	doi = {10.1103/PhysRevB.100.035448},
	number = {3},
	journal = {Physical Review B},
	author = {Bi, Zhen and Yuan, Noah F. Q. and Fu, Liang},
	month = jul,
	year = {2019},
	note = {Publisher: American Physical Society},
	pages = {035448},
}

@article{Kerelsky2019,
  doi = {10.1038/s41586-019-1431-9},
  url = {https://doi.org/10.1038/s41586-019-1431-9},
  year = {2019},
  month = jul,
  publisher = {Springer Science and Business Media {LLC}},
  volume = {572},
  number = {7767},
  pages = {95--100},
  author = {Alexander Kerelsky and Leo J. McGilly and Dante M. Kennes and Lede Xian and Matthew Yankowitz and Shaowen Chen and K. Watanabe and T. Taniguchi and James Hone and Cory Dean and Angel Rubio and Abhay N. Pasupathy},
  title = {Maximized electron interactions at the magic angle in twisted bilayer graphene},
  journal = {Nature}
}

@article{Parker2021,
  doi = {10.1103/physrevlett.127.027601},
  url = {https://doi.org/10.1103/physrevlett.127.027601},
  year = {2021},
  month = jul,
  publisher = {American Physical Society ({APS})},
  volume = {127},
  number = {2},
  author = {Daniel E. Parker and Tomohiro Soejima and Johannes Hauschild and Michael P. Zaletel and Nick Bultinck},
  title = {Strain-Induced Quantum Phase Transitions in Magic-Angle Graphene},
  journal = {Physical Review Letters}
}

@article{Tilak2021,
  doi = {10.1038/s41467-021-24480-3},
  url = {https://doi.org/10.1038/s41467-021-24480-3},
  year = {2021},
  month = jul,
  publisher = {Springer Science and Business Media {LLC}},
  volume = {12},
  number = {1},
  author = {Nikhil Tilak and Xinyuan Lai and Shuang Wu and Zhenyuan Zhang and Mingyu Xu and Raquel de Almeida Ribeiro and Paul C. Canfield and Eva Y. Andrei},
  title = {Flat band carrier confinement in magic-angle twisted bilayer graphene},
  journal = {Nature Communications}
}

@book{Lardner1971,
  doi = {10.3138/9781487585877},
  url = {https://doi.org/10.3138/9781487585877},
  year = {1971},
  month = dec,
  publisher = {University of Toronto Press},
  author = {R.W. Lardner},
  title = {Mathematical Theory of Dislocations and Fracture}
}

@article{burgers1940geometrical,
  title={Geometrical considerations concerning the structural irregularities to be assumed in a crystal},
  author={Burgers, JM},
  journal={Proceedings of the Physical Society (1926-1948)},
  volume={52},
  number={1},
  pages={23},
  year={1940},
  publisher={IOP Publishing}
}

@article{Koshino2019,
  doi = {10.1103/physrevb.100.075416},
  url = {https://doi.org/10.1103/physrevb.100.075416},
  year = {2019},
  month = aug,
  publisher = {American Physical Society ({APS})},
  volume = {100},
  number = {7},
  author = {Mikito Koshino and Young-Woo Son},
  title = {Moir{\'{e}} phonons in twisted bilayer graphene},
  journal = {Physical Review B}
}

@article{Beechem2014,
  doi = {10.1021/nn405999z},
  url = {https://doi.org/10.1021/nn405999z},
  year = {2014},
  month = jan,
  publisher = {American Chemical Society ({ACS})},
  volume = {8},
  number = {2},
  pages = {1655--1663},
  author = {Thomas E. Beechem and Taisuke Ohta and Bogdan Diaconescu and Jeremy T. Robinson},
  title = {Rotational Disorder in Twisted Bilayer Graphene},
  journal = {{ACS} Nano}
}
\newpage
\appendix
\section*{Supplementary information}
\section{AFM comparison}\label{sec:AFM}
\begin{figure}[!ht]
\centering
\includegraphics[width=0.9\columnwidth]{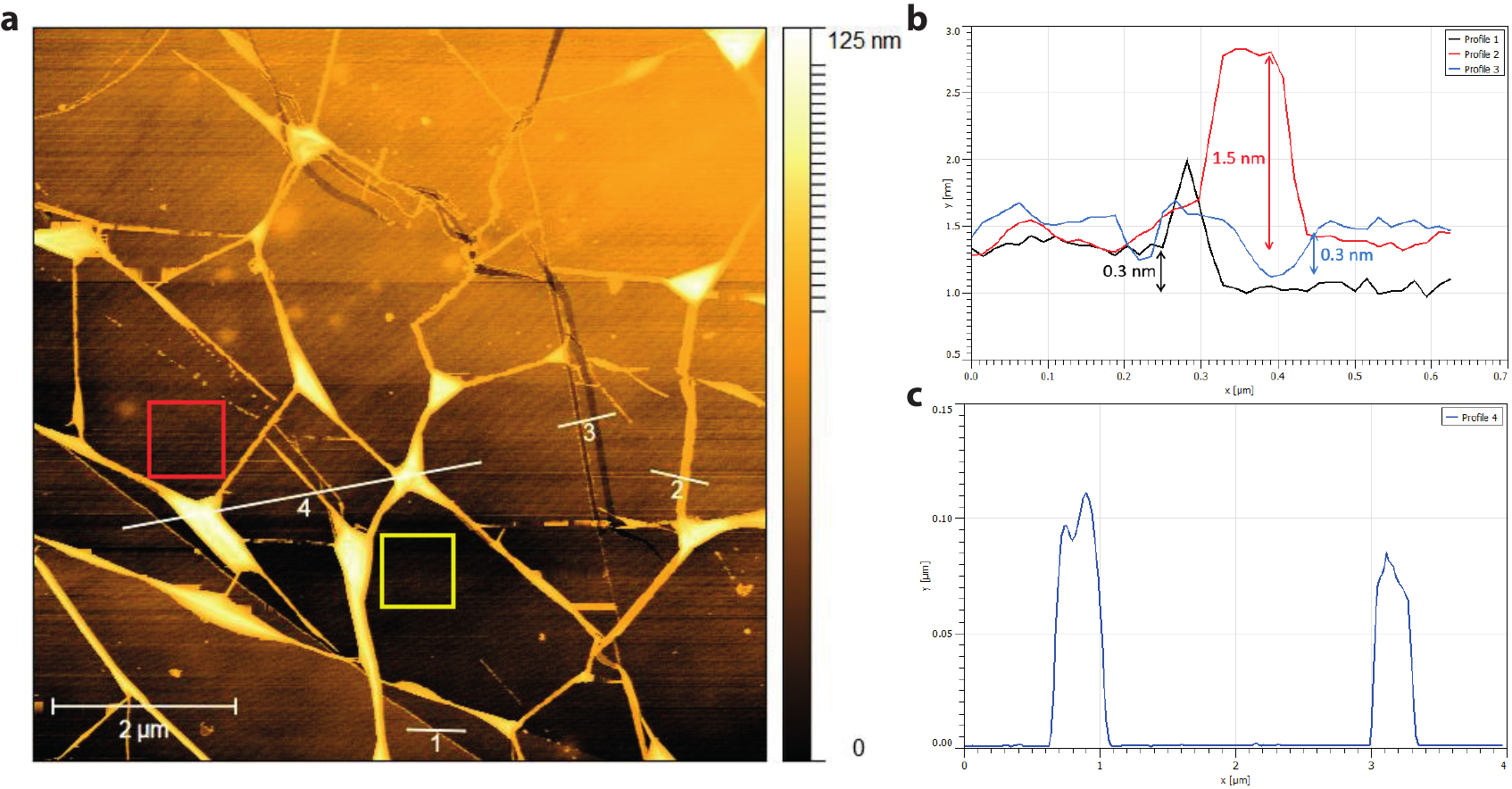}
\caption{\subf{a} Atomic Force Microscopy overview  of sample area. Locations of line profiles and detailed topographies in \figref{AFM_dislocations} are indicated.
\subf{b,c} Line cuts along the cuts indicated in \subref{a}.}
\label{fig:AFM}
\end{figure}

\begin{figure}[!ht]
\centering
\includegraphics[width=0.4\columnwidth]{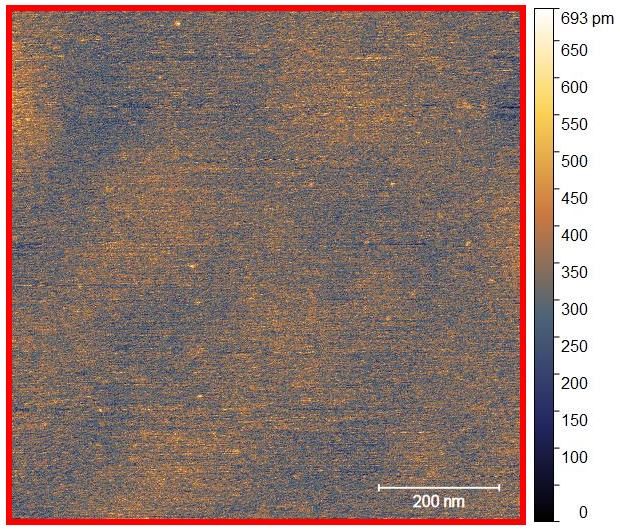}
\includegraphics[width=0.4\columnwidth]{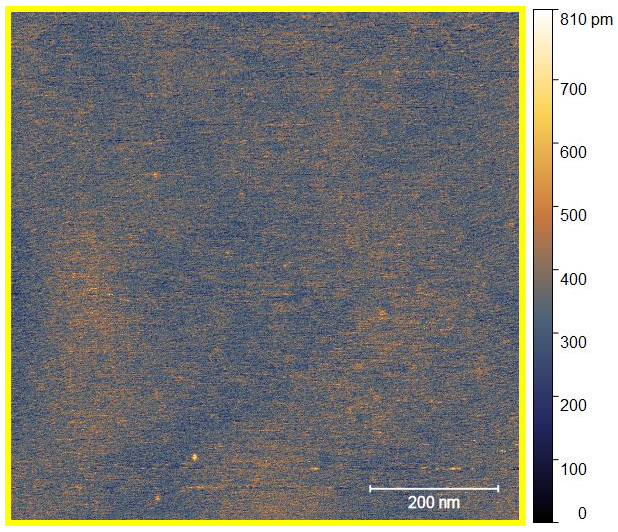}
\caption{\subf{a} Atomic Force Microscopy of the dislocation area in \figref[a]{dislocation2}. Area is indicated in red in \figref[a]{AFM}.
\subf{b} Atomic Force Microscopy of the dislocation area in \figref[b]{dislocation2}. Area is indicated in yellow in \figref[a]{AFM}.}
\label{fig:AFM_dislocations}
\end{figure}
To further characterize the surface properties of the sample, an AFM (JPK, NanoWizard 3) measurement was performed in AC tapping mode following the LEEM measurements. Predominantly, the results show a very flat and clean graphene surface between folds, indicating annealing at 500 $^{\circ}$C in UHV had successfully removed the polymer residue left on the surface.

In profile 1, the terrace height sees a difference of 0.3\,nm, demonstrating the graphene layer count goes down by one at this location. This corresponds to the layer counts extracted from the LEEM spectra.

Profile 2 to 4 shows 3 different kinds of defects in the bilayer graphene region. The ridge at location 2 seems to be a neat folding of both the bilayer graphene flake (1.5\,nm in height, four layers of graphene), whereas profile 4 shows wrinkles that are up to 120\,nm tall. This is also reflected by the distinct patterns in the LEEM bright field overview image, respectively. While the wrinkles merely appear black, the ridge resembles more like a unique layer count domain. Profile 3 shows two tears within the one layer of graphene, corresponding nicely to the defect region observed in LEEM where monolayer graphene shines through.

The zoomed-in small scale measurements marked by the red and yellow box shows the topography on top of two dislocations observed in LEEM.  As shown in \figref{AFM_dislocations}, no distinctive feature was observed at either dislocation. The topography, however, shows an exceptionally flat surface with a height variation (peak-to-peak) of less than 1\,nm.
\FloatBarrier

\section{Supplementary notes on Adaptive GPA}\label{sec:SIextendedGPA}
Regular GPA is limited in the wave vector deviations (with respect to the reference wave vector) it can measure, due to the limitations in spectral leakage. This is no problem when applied to atomic lattices, as the expected deviations are very small there. However, due to the moir\'e magnification of small lattice distortions, it does become a limiting factor when applying GPA to small twist angle moir\'e lattices.

To overcome this limitation, we extended the GPA algorithm to use adaptive reference wave vectors, based on the combination of two ideas and related to earlier work in laser fringe analysis~\cite{Kemao2007}:
First, a GPA phase calculated with respect to one reference vector can always be converted to the GPA phase with respect to another reference vector by adding a phase corresponding to the phase difference between the reference vectors.
Second, a larger lock-in amplitude corresponds to a better fit between the reference vector and the data.

The adaptive GPA algorithm therefore works as follows:
The spatial lock-in signal is calculated for a grid of wave vectors around a base reference vector, converting the GPA phase to reference the base reference vector every time. For each pixel, the spatial lock-in signal with the highest amplitude is selected as the final signal.

It was realized that to deduce the deformation properties, reconstruction to a globally consistent phase (requiring 2D phase unwrapping), as reported previously~\cite{benschop2020mapping}, is not strictly needed, making it possible to circumvent the problems associated with 2D phase unwrapping. 
Instead, the gradient of each GPA phase was calculated, requiring only local 1D phase unwrapping (i.e. assuming the derivative of the phase in both the $x$ and $y$ direction will never be more than $\pi$ per pixel, an assumption in practice always met). Subsequently, these three GPA gradients are converted to the displacement gradient tensor (in real space coordinates), estimating the transformation via weighted least squares, using the local spatial lock-in amplitudes as weights.

As an added benefit, this entire procedure is local, i.e. not depending on pixels beyond nearest neighbors in any way except for the initial Gaussian convolution in determining the GPA.  This reduces the effect of artefacts in the image to a minimum local area around each artefact (where for in the 2D phase unwrapping they have a global influence on the phases).

However, when the gradient is computed based on phase values stemming from two different GPA reference vectors, i.e. at the edge of their valid/optimal regions, artefacts appear due to their relatively large absolute error. To prevent this, the local gradient of the phase with the highest lock-in magnitude is stored alongside the lock-in signal itself in the GPA algorithm. This way, the gradient is calculated based on a single reference phase, propagating only the much smaller relative/derivative error between the two signals instead of the absolute error.

As mentioned in the main text, even adaptive GPA has its limits. In particular, too large deviations from the base reference vector can not be resolved correctly, causing an erroneous, lower, extracted deviation, as is visible in the lower right of \figref[e]{distortion} in the main text). As the deformation becomes too large, e.g. towards the folds in the TBG, the highest lock-in amplitude will occur at a different moir\'e peak or at the near-zero components of the fourier transform, causing an incorrect value to be extracted.\\

\FloatBarrier
\section{Supplementary note on the decomposition of the displacement field.}
Kerelsky et al.~\cite{Kerelsky2019} use the following idea to extract twist angle $\theta_T$, strain magnitude $\epsilon$ and strain direction $\theta_s$ from reciprocal moir\'e lattice vectors $K_{is}$: These difference vectors of the constituting atomic lattices are written in terms of a rotated and a strained lattice vector each:
$$K_{is} = k_{ir} - k_{is} = R(\theta_T)k_i-S(\theta_s,\epsilon)k_i$$
where $k_i$ are the original lattice vectors. Kerelsky et al. assume $k_0$ to be along the $x$-axis, and get around this by taking amplitudes, discarding any global rotation. Here, we do however introduce that global rotation, by a multiplication with $R(\xi)$:
$$K_{is} = \left(R(\theta_T)-S(\theta_s,\epsilon)\right)R(\xi)k_i$$
Eihter of these expressions can, and indeed by Kerelsky et al. is, numerically fitted to the found amplitudes or $k$-vectors for each triangle. 
However, from GPA analysis we most naturally obtain a Jacobian transformation $J_{ac}$ of the moiré $k$-vectors with respect to some specific set of reference vectors with predefined strain and rotations:
$$K_{is} = (J+I)K_{i0} = J_{ac}K_{i0} = J_{ac} \left(R(\theta_{T0})-S(\theta_{s0},\epsilon_{0})\right)R(\xi_0)k_i := J_{ac}A_0R(\xi_0)k_i$$
Note that we can force $\epsilon_0=0 \rightarrow S(\theta_{s0},\epsilon_{0}) = I$.\\
This simplifies to:
\begin{align*}
J_{ac}A_0R(\xi_0)k_i &= \left(R(\theta_T)-S(\theta_s,\epsilon)\right)R(\xi)k_i\\
\intertext{The linear transformation is uniquely described by its effect on two points in $k$-space, so their matrix representations should be equal:}
J_{ac}A_0R(\xi_0) &= \left(R(\theta_T)-S(\theta_s,\epsilon)\right)R(\xi)\\
J_{ac}A_0 &= \left(R(\theta_T)-S(\theta_s,\epsilon)\right)R(\xi-\xi_0)\\
\end{align*}
The left hand side is a known quantity at each position, the right hand side remains to be numerically fitted or extracted. This is implemented in \texttt{pyGPA} using \texttt{scipy.optimize} and \texttt{numba} to just-in-time compile the fitting code~\cite{pyGPA,numba}.

Alternatively, we could formulate a symmetric expression with two strains, but without allowing for further joint rotation of the lattices:
$$K_{i0} = \left(R(\theta_{T0}/2)-R(-\theta_{T0}/2)\right)R(\xi_0)k_i := B_0R(\xi_0)k_i$$
\begin{align*}
J_{ac}B_0R(\xi_0)k_i &= \left(S(\theta_b,\epsilon_b)R(\theta_T/2)-S(\theta_a,\epsilon_a)R(-\theta_{T}/2)\right)R(\xi_0)k_i\\
J_{ac}B_0 &= \left(S(\theta_b,\epsilon_b)R(\theta_T/2)-S(\theta_a,\epsilon_a)R(-\theta_{T}/2)\right)\\
\end{align*}

\section{LEEM stitching}
\begin{figure}[!ht]
\centering
\includegraphics[width=0.8\columnwidth]{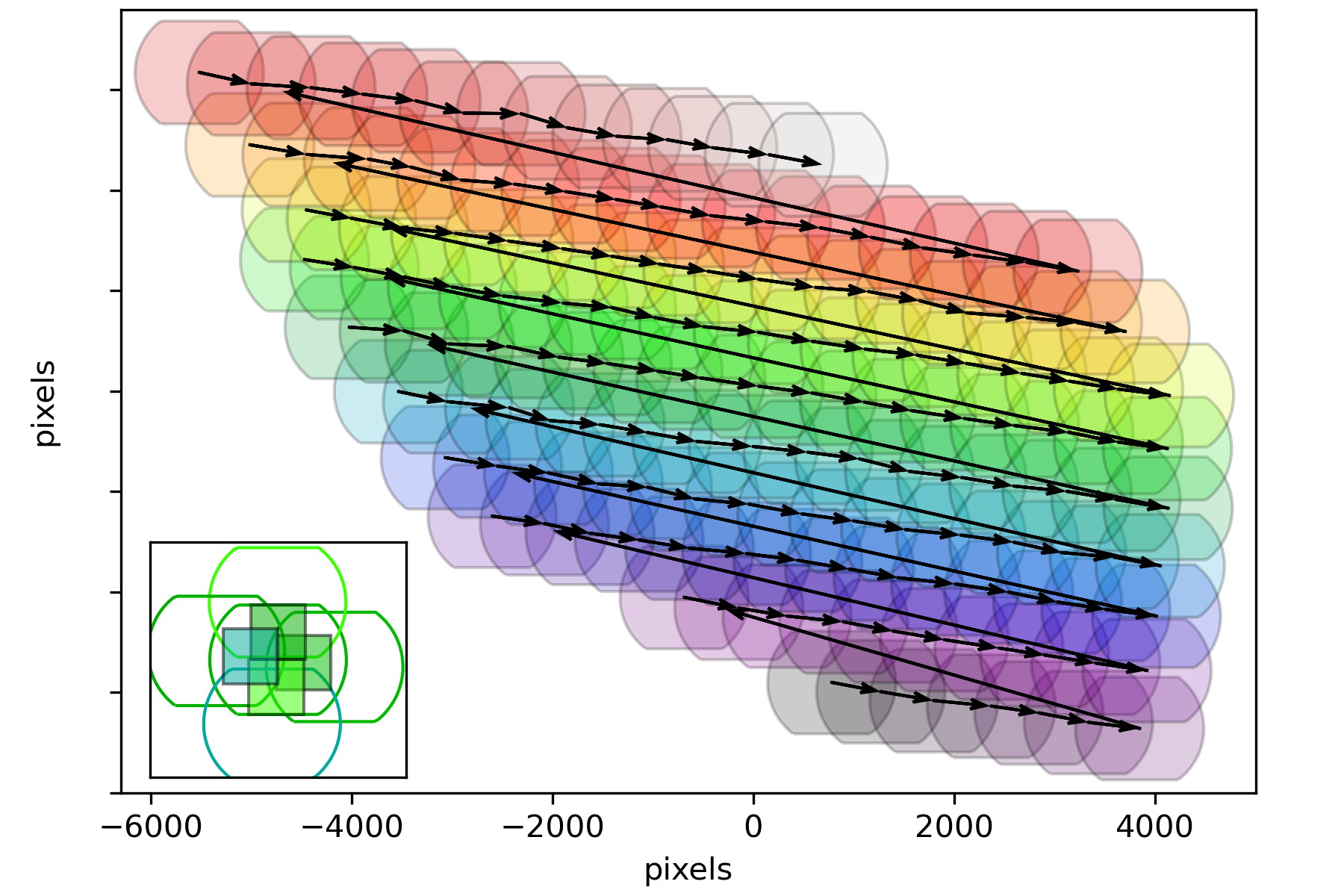}
\caption{Illustration of the sample stage scanning for stitched overview images. Black arrows indicate the direction of stage movement. \textbf{inset,} Illustration of the square overlapping regions of neighboring images used to determine relative positions.}
\label{fig:SIstitching}
\end{figure}
To achieve stitching of images without inducing any additional deformation, a custom stitching algorithm tailored towards such LEEM data, was developed, working as follows:

To compensate sample stage inaccuracy, nearest neighbor (by sample stage coordinates) images are compared, finding their relative positions by cross-correlation. Using an iterative procedure, calculating cross correlations of overlapping areas at each step, the absolute positions of all images are found. 
Images are then combined in a weighted fashion, with the weight sloping to zero at the edges of each image, to smooth out any mismatch due to residual image warping.
The full stitching algorithm is implemented in \texttt{Python}, available as a \texttt{Jupyter Notebook}\cite{LEEM-analysis}. 

It is
designed for use with ESCHER LEEM images. For those images, their
positions are known approximately in terms of \emph{stage coordinates},
i.e.~the positions as reported by the sensors in the sample stage. It
should however generalize to any set of overlapping images where
relative positions of the images are known in some coordinate system
which can approximately be transformed to coordinates in terms of pixels
by an affine transformation (rotation, translation, mirroring).

\begin{figure}[!ht]
\centering
\includegraphics[width=0.8\columnwidth]{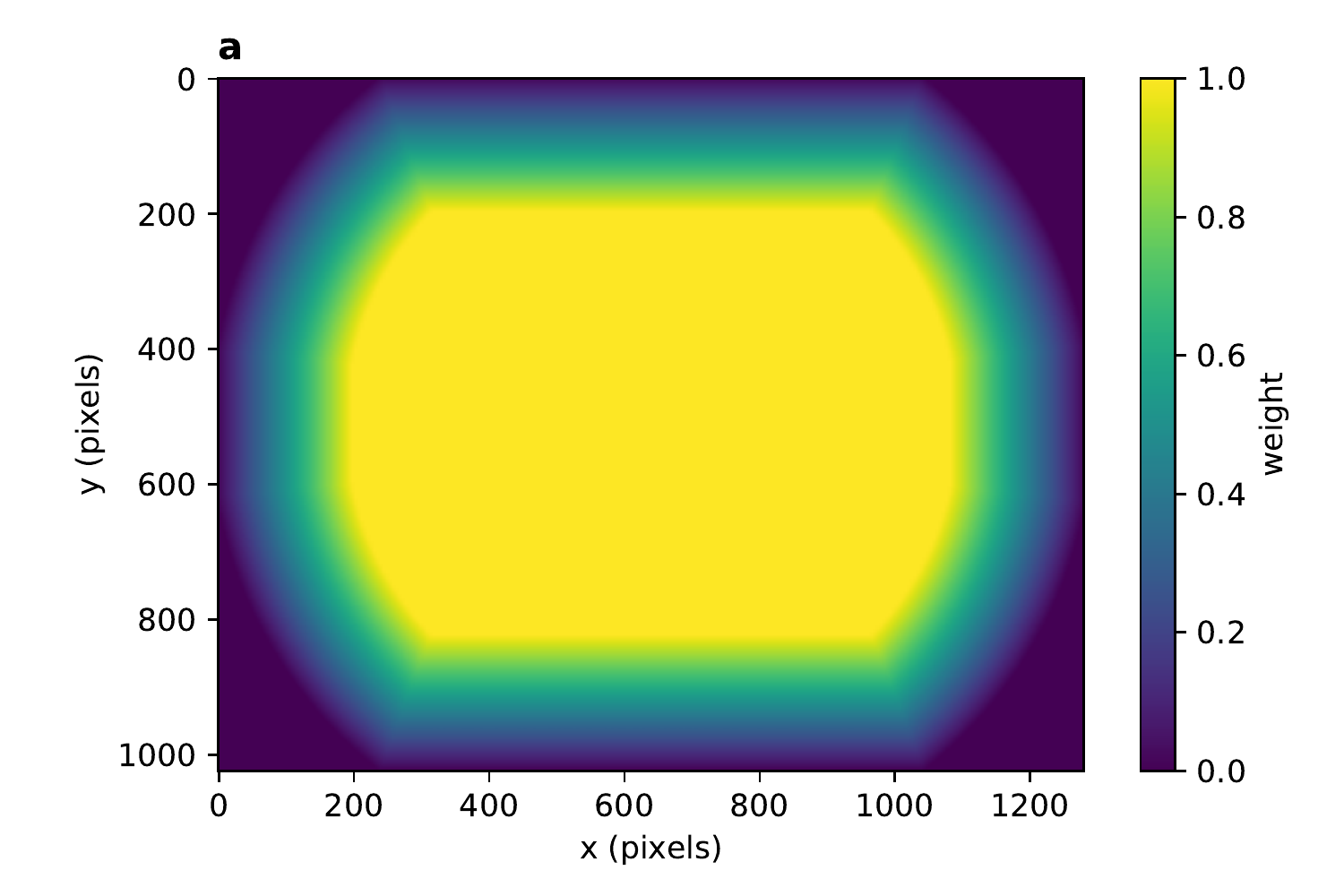}
\caption{\subf{a} Weight mask used to merge images. A linear slope of the weight towards the edges of the round microchannel plate detector is used to smoothly merge images.}
\label{fig:SIweightmask}
\end{figure}

The algorithm consists of the following steps:

\begin{enumerate}
\def\labelenumi{\arabic{enumi}.}
\item
  Using the stage coordinates for each image, obtain a nearest neighbour
  graph with the nearest \texttt{n\_neighbors} neighbouring images for
  each image.
\item
  Obtain an initial guess for the transformation matrix between stage
  coordinates and pixel coordinates, by one of the following options:
  \begin{enumerate}
  \def\labelenumii{\arabic{enumii}.}
\item
  Copying a known transformation matrix from an earlier run of a
  comparable dataset.
\item
  Manually overlaying some nearest neighbor images from the center of
  the dataset, either refining the estimate, or making a new estimate
  for an unknown dataset
  \end{enumerate}
\item
  Calculate an initial estimate of the pixel coordinates of the images
  by applying the corresponding transformation to the stage coordinates
\item
  Apply a gaussian filter with width \texttt{sigma} to the original
  dataset and apply a magnitude sobel filter. Optionally scale down the
  images by an integer factor \texttt{z} in both directions to be able
  to reduce \texttt{fftsize} by the same factor, without reducing the
  sample area compared.
\item
  Iterate the following steps until the calculated image positions have
  converged to within \texttt{sigma}:
  \begin{enumerate}
  \def\labelenumii{\arabic{enumii}.}
\item
  Obtain a nearest neighbour graph with per image the nearest
  \texttt{n\_neighbors} neighbouring images from the current estimate of
  the pixel coordinates and calculate the difference vectors between
  each pair of nearest neighbours.
\item
  For each pair of neighboring images:

  \begin{enumerate}
  \def\labelenumii{\arabic{enumii}.}
  \item
    Calculate the cross-correlation between areas estimated to be in the
    center of the overlap of size \texttt{fftsize*fftsize} of the
    filtered data. If the estimated area is outside the valid area of
    the image defined by \texttt{mask}/\texttt{radius}, take an area as
    close to the intended area but still within the valid area as
    possible.
  \item
    Find the location of the maximum in the cross-correlation. This
    corresponds to the correction to the estimate of the difference
    vector between the corresponding image position pair.
  \item
    Calculate the weight of the match by dividing the maximum in the
    cross-correlation by the square root of the maximum of the
    auto-correlations.
  \end{enumerate}
\item
  Compute a new estimate of the difference vectors by adding the found
  corrections. Reconvert to a new estimate of pixel coordinates by
  minimizing the squared error in the system of equations for the
  positions, weighing by modified weights, either:

  \begin{enumerate}
  \def\labelenumii{\arabic{enumii}.}
  \item
    \(w_{mod}= w - w_\textit{min}\) for \(w> w_\textit{min}\), \(w=0\) else, with
    $w_\textit{min}$ the maximum lower bound such that the graph of nearest
    neighbours with non-zero weights is still connected
  \item
    Only use the `maximum spanning tree' of weights, i.e.~minus the
    minimum spanning tree of minus the weights, such that only the \(n\)
    best matches are used.
  \end{enumerate}
 \end{enumerate}
\item
  (Optional) Refine the estimate of the transformation matrix, using all
  estimated difference vectors with a weight better than \(w_{min est}\)
  and restart from step 3.
\item
  Repeat step 4. and 5. until \texttt{sigma} is satisfactory small.
  Optional repeat a final time with the original data if the signal to
  noise of the original data permits.
\item
  Select only the images for stitching where the average of the used
  weights (i.e.~where \(w > w_{min}\)) is larger than \(q_\textit{thresh}\) for
  an appropriate value of \(q_\textit{thresh}\).
\item
  (Optional) For those images, match the intensities by calculating the
  intensity ratios between the overlap areas of size
  \texttt{fftsize*fftsize} and perform a global optimization.
\item
  Define a weighting mask, 1 in the center and sloping linearly to zero
  at the edges of the valid region, over a width of \texttt{bandwidth}
  pixels, as illustrated in \figref{SIweightmask}.
\item
  Per block of output \texttt{blocksize*blocksize}, select all images
  that have overlap with the particular output block, multiply each by
  the weighting mask and shift each image appropriately. Divide by an
  equivalently shifted stack of weighting masks. As such information at
  the center of images gets prioritized, and transitions get smoothed.
\end{enumerate}

\subsection{Considerations}\label{considerations}

For square grids with a decent amount of overlap, it makes sense to put
\texttt{n\_neighbors} to 5 (including the image itself), however, for
larger overlaps or datasets where an extra dimension is available (such
as landing energy), it can be appropriate to increase the number of
nearest neighbors to which each image is matched.

Parameters and intermediate results of the iteration are saved in an
\texttt{xarray} and saved to disk for reproducibility.

\subsection{Parallelization}\label{parallelization}

Using \href{https://dask.org}{\texttt{dask}}, the following steps are
parallelized:

\begin{itemize}
\item
  step 5B , where each pair of images can be treated independently. In
  practice parallelization is performed over blocks of subsequent images
  with their nearest neighbours. This could be improved upon in two
  ways: firstly by treating each pair only once, and secondly by making
  a nicer selection of blocks of images located close to each other in
  the nearest neighbor graph. This would most likely require another (smarter) data
  structure than the nearest neighbour indexing matrix used now.
\item
  Step 6 is quite analogous to 5B and is parallelized
  similarly.
\item
  Step 11 is parallelized on a per block basis. To optimize memory
  usage, results are directly streamed to a \texttt{zarr} array on disk.
\item
  The minimizations are parallelized by \texttt{scipy} natively.
\end{itemize}

\FloatBarrier
\section{Additional images/crops}\label{sec:SIB}

\begin{figure}[!ht]
\centering
\includegraphics[width=0.95\columnwidth]{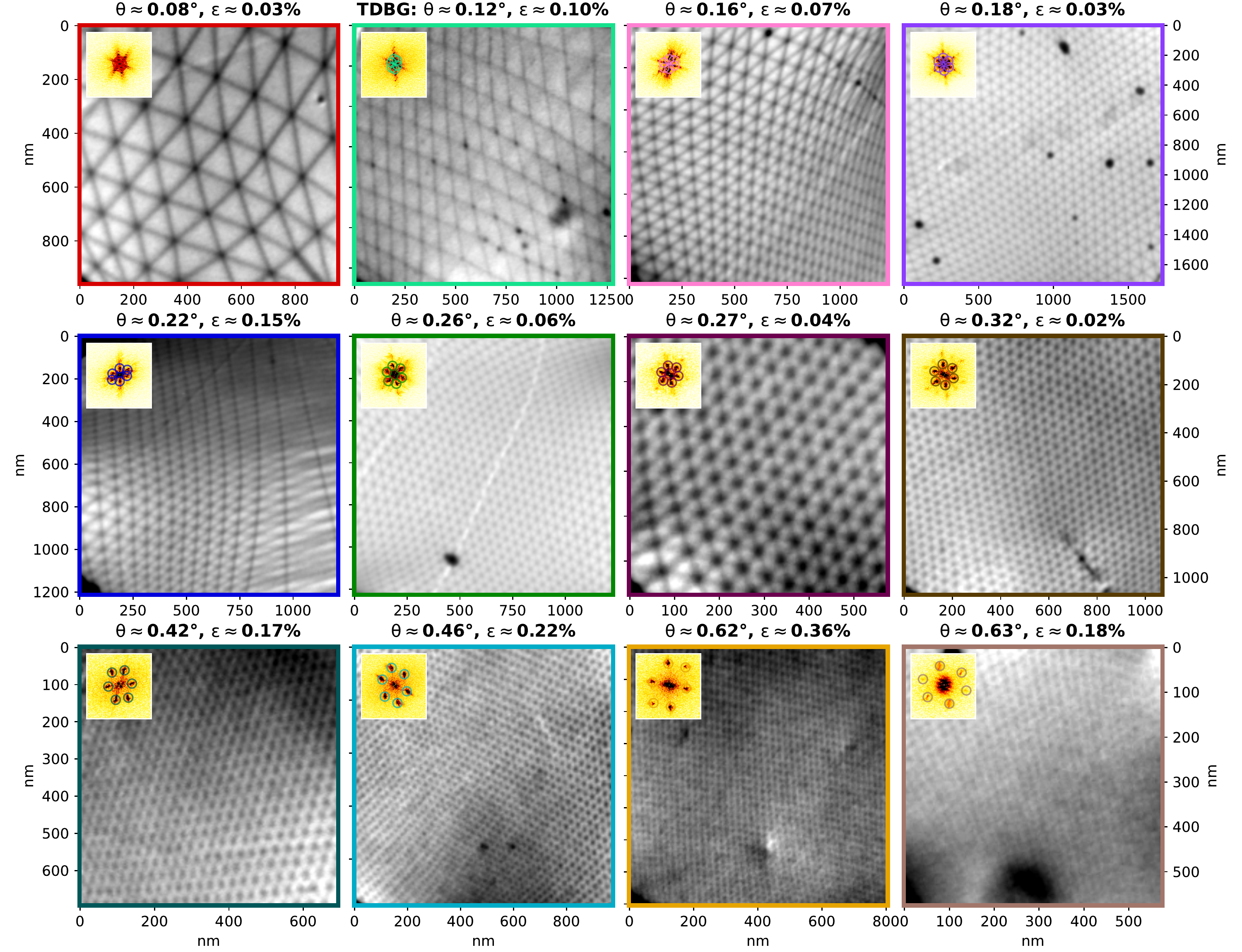}
\caption{A wider range of images found in the sample from the main text as used to determine the histograms of twist angles and strain in Figure 2 of the main text. Insets show FFT's with the detected moiré peaks}
\label{fig:moretwistangles}
\end{figure}

\begin{figure}[!ht]
\centering
\includegraphics[width=\columnwidth]{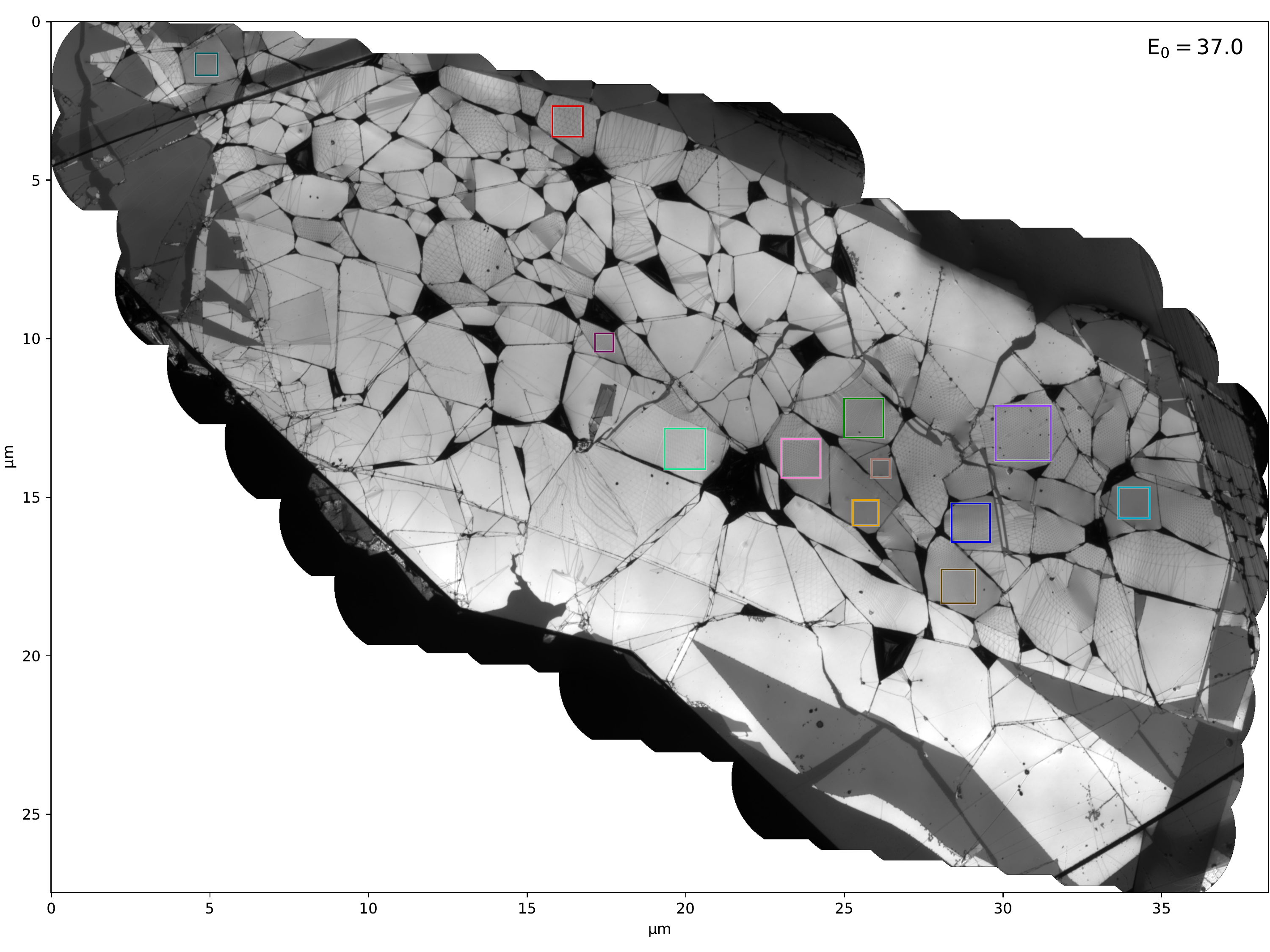}
\caption{Locations of the crops in \figref{moretwistangles} indicated in the full overview (data is the same as \figref[d]{overview} of the main text).}
\label{fig:moretwistanglesloc}
\end{figure}

\begin{figure}[!ht]
\centering
\includegraphics[width=0.9\columnwidth]{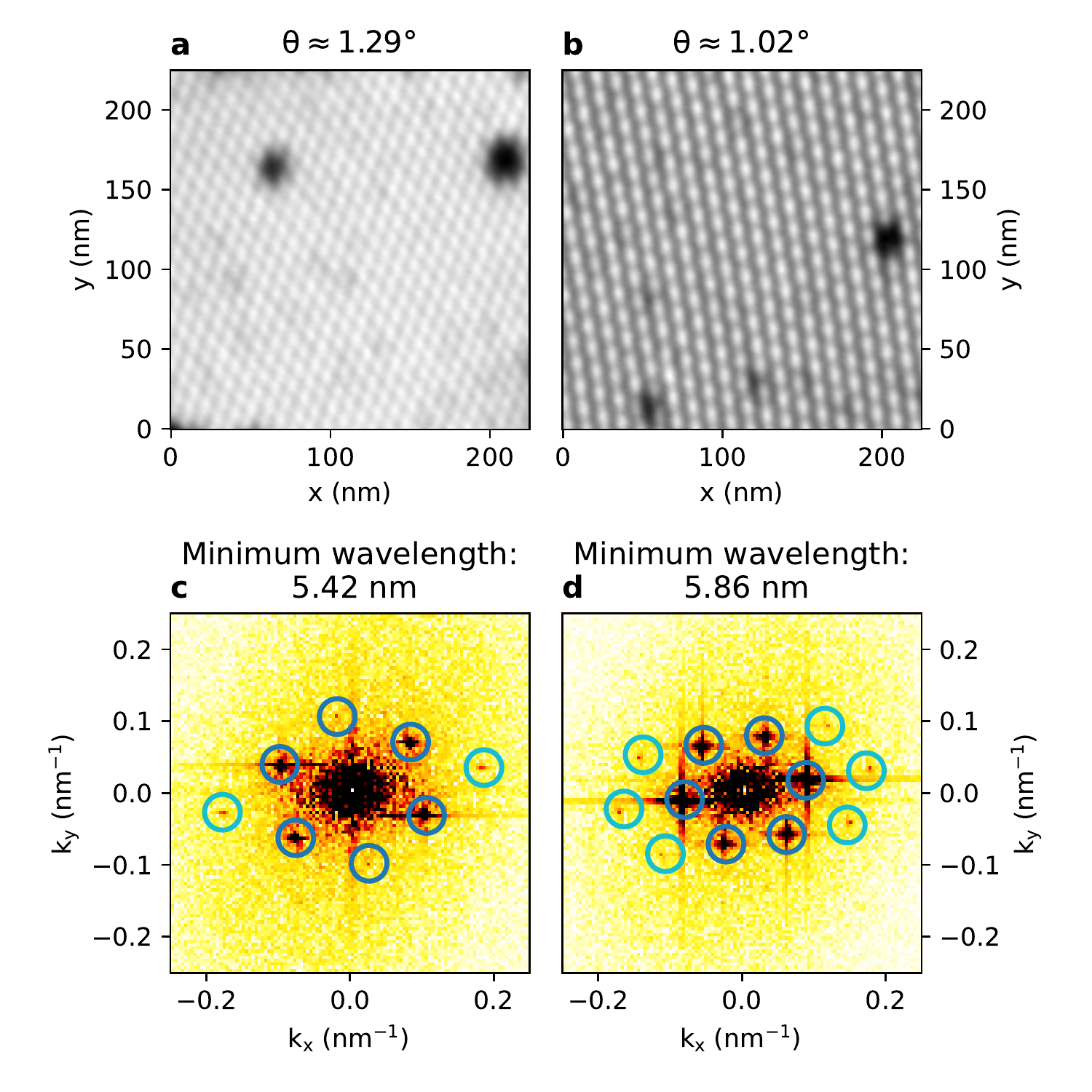}
\caption{\subf{a,b} Images of 2-on-2 layer twisted graphene at $E_0=17.0\,\text{eV}$, with twist angles of respectively $\theta \approx 1.29^\circ$ and $\theta \approx 1.02^\circ$
\subf{c,d} FFTs of a,b with Bragg peaks corresponding to the moiré pattern indicated in blue. Higher order moiré peaks are also visible (indicated in cyan), corresponding to minimum detectable wavelengths of less than 6\,nm.}
\label{fig:resolution}
\end{figure}

\FloatBarrier
\section{Supplementary figures dislocations}\label{sec:SI_disl}

\begin{figure}[!ht]
\centering
\includegraphics[width=0.95\columnwidth]{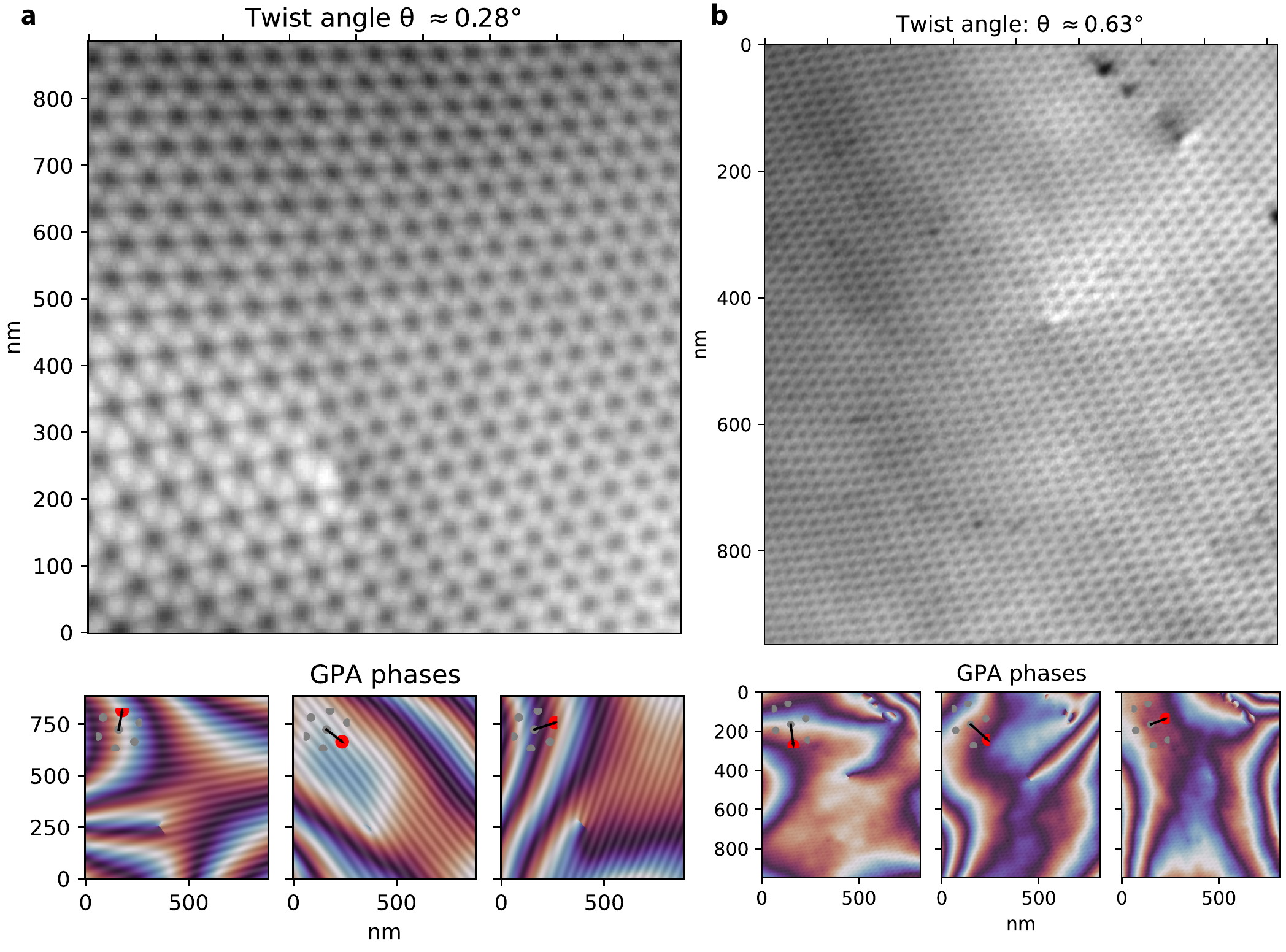}
\caption{\subf{a} Additional edge dislocation found on the sample at a lower twist angle. 
\subf{b} Larger area around edge dislocation in the main text from the main text.
In both case GPA phases are also displayed.}
\label{fig:dislocation2}
\end{figure}

\begin{figure}[!ht]
\centering
\includegraphics[width=0.9\columnwidth]{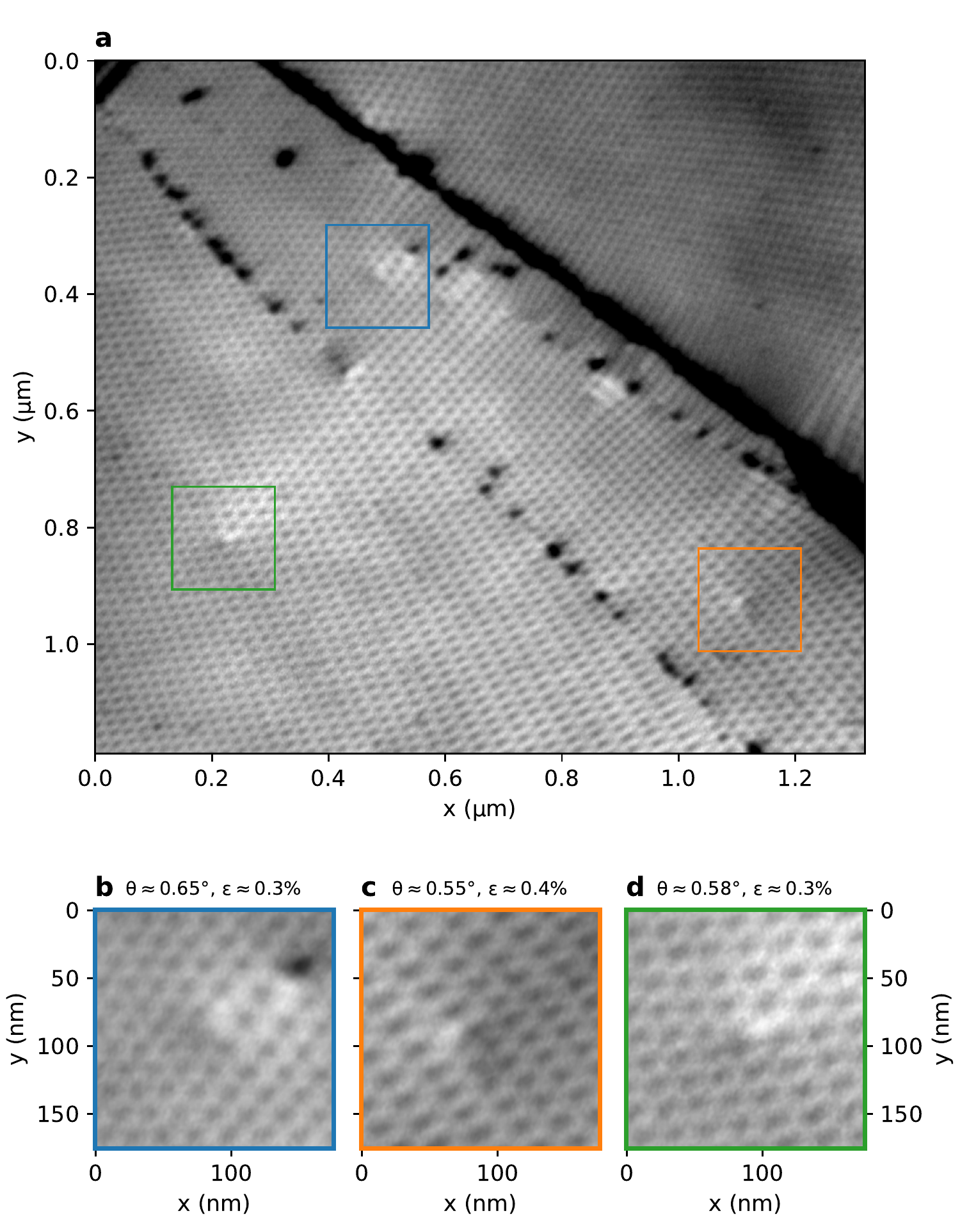}
\caption{More dislocations in the vicinity of the dislocation shown in the main text.
\subf{d} corresponds to the dislocation in the main text. $\theta$ as extracted from the shown area here is a bit lower as unit cell area tends to be a bit larger near the dislocation.}
\label{fig:dislocation3}
\end{figure}

\begin{figure}[!ht]
\centering
\includegraphics[width=\columnwidth]{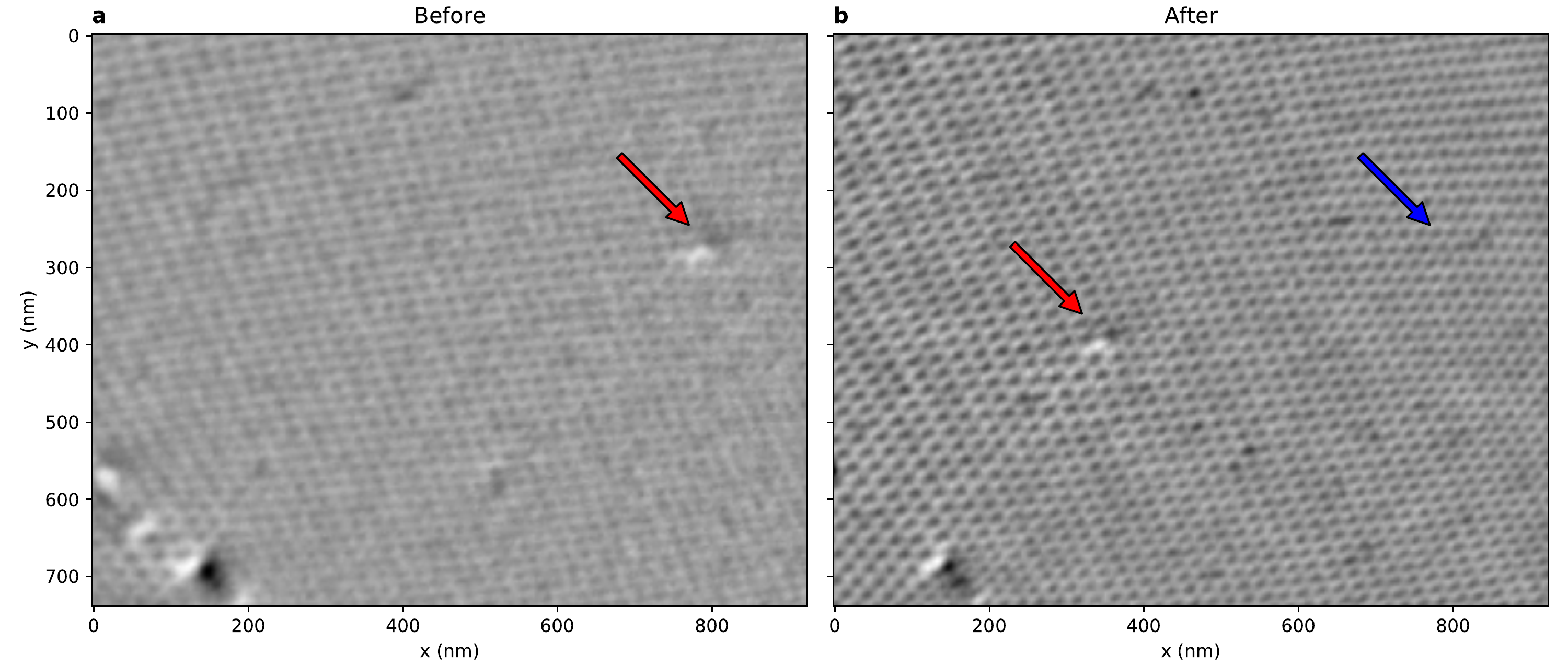}
\includegraphics[width=\columnwidth]{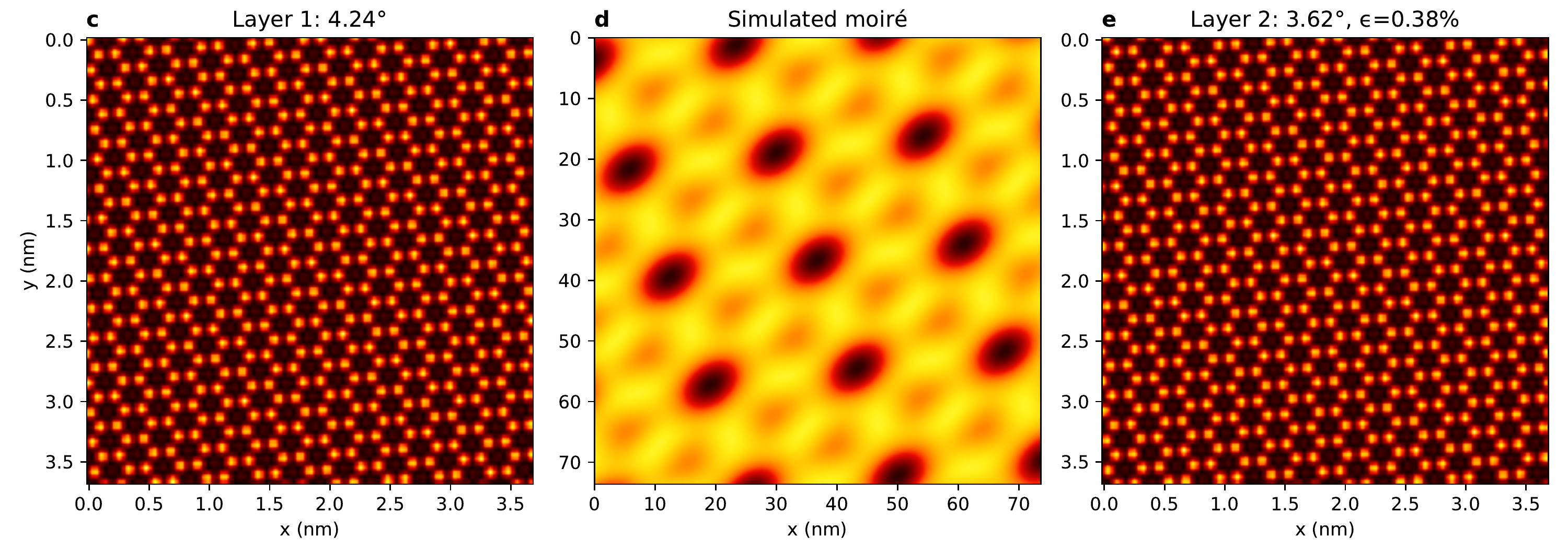}
\caption{\textbf{Movement of dislocation.} \subf{a} Dislocation in its original location, indicated by red arrow. 
\subf{b} Image of the same area as in \subf{a}, but imaged two days later. The dislocation has moved, as indicated by the red arrow. The former location is indicated with a blue arrow.
\subf{c-e} Rendering of the individual atomic lattices and the resulting moiré lattice from the extracted lattice parameters, showing the atomic lattice directions.}
\label{fig:dislocation_move}
\end{figure}

\FloatBarrier
\section{Dynamics}
Full movie showing larger Field of View compared to \figref{dynamics}, in real space data, difference data and GPA-extracted displacement field.

\end{document}